\documentclass[aps,pre,reprint,groupedaddress,onecolumn,longbibliography]{revtex4-2}

\usepackage{hyperref}
\usepackage{graphicx}
\usepackage{dcolumn}
\usepackage{bm}
\usepackage{subcaption}
\usepackage{amssymb}
\usepackage{amsmath}
\usepackage{xcolor}

\begin{document}


\title{Turbulent heat transfer enhancement by compliant walls}


\author{Morie Koseki}
\author{Marco Edoardo Rosti}
\email[E-mail for correspondence: ]{marco.rosti@oist.jp}
\affiliation{Complex Fluids and Flows Unit, Okinawa Institute of Science and Technology Graduate University, 1919-1 Tancha, Onna-son, Okinawa 904-0495, Japan.}


\date{\today}

\begin{abstract}
This study investigates the effect of compliant walls on the turbulent heat transfer in channel flows over viscous-hyperelastic walls. We perform Direct Numerical Simulations, fully resolving the mutual fluid-structure interactions between the turbulent flow and the compliant walls, varying the wall elasticity and the thermal diffusivity in a fully turbulent condition. We show that the compliant wall leads to an increase not only of the momentum transfer but also of the heat transfer. Since the compliant wall can dynamically move, in the near-wall region heat flux is mostly transferred via turbulent convection rather than diffusion, as typically found with rigid walls. Thus, the heat flux can be controlled not only by varying the thermal diffusivity, but also by changing the transverse modulus of elasticity which governs the wall-normal velocity fluctuations and consequently the temperature ones. Finally, we show that the physical mechanism controlling these modifications are the sweep and ejection events.
\end{abstract}

\maketitle

\section{Introduction}
Compliant surfaces are walls that deform in response to the instantaneous stresses acting onto them according to an elastic response. Here we focus on elastic layers which interact with a turbulent flow, with a complex fluid-structure interaction that deforms the boundary between the two phases in time and space. Although the complexity of the problem, the effect of compliant walls on a turbulent flow has drawn significantly attention in both the scientific community and industry as a possible and cheap way to alter the flow, with no additional energy required \cite{riley1988complaint, carpenter1993optimization, nagy2022effect, semenov1991conditions, carpenter2000hydrodynamics, gad2002compliant, nisewanger1964flow}.

The first studies on the topic have been theoretically, mostly focusing on the feasibility of compliant walls to achieve a delay in transition and an attenuation of turbulence \cite{kumaran2021stability}. However, it has been shown that wall compliance can lead to both flow stabilization and de-stabilization \cite{benjamin1960effects, benjamin1963threefold, landahl1962stability, carpenter1985hydrodynamic, carpenter1986hydrodynamic}: the wall compliance stabilizes the Tollmien-Schlichting instability, resulting in transition delay \cite{rotenberry1990effect, rotenberry1992finite, davies1997numerical}, while the wall compliance de-stabilizes a flow-induced surface instability, which can lead to two  new instabilities, the traveling-wave flatter and static divergence \cite{davies1997instabilities}. More recent works have extended the theoretical analysis to the interaction with turbulent flows, showing that pressure instabilities lead to larger amplitudes of the surface wave \cite{duncan1986response}, which the effect of the different parameters being investigated in Ref.~\cite{benschop2019deformation}.

Experimental investigations available in the literature mostly focused on the coupling motion between compliant walls and turbulent flows when varying the level of wall elasticity. It has been shown that, when the wall motions are smaller than few wall units, the system is the so-called one-way coupled state, exhibiting traveling waves that correlate strongly with turbulent pressure fluctuations \citep{zhang2015integrating, zhang2017deformation}. In the two-way coupled regime, when deformations exceed several wall units, spanwise- and streamwise-aligned waves emerge that alter the near-wall momentum transport \citep{wang2020interaction, greidanus2022response}.

Numerical simulations have lagged behind the theoretical analysis and experiments, due to the high complexity of the problem. While early simulations used simplified mass-spring-damper models to model the compliant walls \cite{endo2002direct, kim2014space}, only recently direct numerical simulations have been able to fully resolve the fluid-structure interaction in the presence of a turbulent flow. These new results have shown that the wall modifications lead to significant drag increase due to the enhanced fluctuations in the fluid driven by the wall motion pumping momentum in the flow \cite{Rosti_Brandt_2017, Ardekani_Rosti_Brandt_2019, Esteghamatian2022spatiotemporal, Koseki_Aswathy_Rosti_2025}. The drag enhancement is accompanied by the modification of the coherent structures present in the turbulent flow; in particular, the streamwise structures are annihilated and replaced by spanwise-oriented ones \cite{Rosti_Brandt_2017}, which enhance the momentum transfer. As the wall compliance increases, ejections prominently contribute to the turbulent production rather than sweep \cite{Ardekani_Rosti_Brandt_2019}, driven by the negative vorticity lifts up with low-speed fluid from the compliant wall \cite{Esteghamatian2022spatiotemporal}. Finally, it has been shown that the effects described above cannot be fully ascribed to the roughness of the deformed interface, with the wall elasticity inducing peculiar alteration to the flow, thus making the simulations of the coupled wall-fluid motions essential to properly describe the flow at hand \cite{Koseki_Aswathy_Rosti_2025}.

No studies have investigate the effect of compliant walls on the turbulent heat transfer, which is the topic of the present investigation. However, both momentum and heat transfers have been studied in the field of rough and porous walls. In particular, it is now well known that these complex walls not only enhance the turbulent moment transfer but also the heat transfer, and more in general the mixing of passive scalar. While specific properties of wall roughness can cause various alterations, overall roughness enhances drag and induces wall-normal perturbations that modify the intensity of the turbulent structure populating the flow \cite{jimenez2004turbulent, kadivar2021review, chung2021predicting}. As a consequence, turbulent heat transfer is also strengthened, through increased ejection and sweep events \cite{KADIVAR2025turbulent}. Similarly, wall permeability also affects the near-wall turbulent structures through the reduction of the wall-blocking effect \cite{manes2009turbulence, manes2011turbulent, rosti2018turbulent, okazaki2020turbulence}. The typical streamwise turbulent structures are disturbed when increasing the permeability by Kelvin-Helmholtz rollers appearing above the porous interface \cite{Breugem2006The, SUGA2010effects, Suga2011vortex, kuwata_lattice_2016}, similarly to what observed in canopy flows \cite{monti_olivieri_rosti_2023a, lohrer_frohlich_2023a, rota2024dynamics, lohrer_frohlich_2025a, marchioli_rosti_verhille_2025a}. Turbulent heat transfer is enhanced also above isotropic porous-walls due to the larger vortical structures developed \cite{chandesris2013direct, JOUYBARI2021investigation}. The effects of anisotropic permeability on turbulent heat flux are consistent with those observed in the turbulent-flow fields \cite{rosti_brandt_pinelli_2018a, nishiyama2020direct} with the spanwise permeability strongly altering the thermal fields \cite{kuwata_suga_2017_direct, suga2018anisotropic}.

Since it has been shown extensively that surface roughness and permeability are able to enhance turbulent heat transfer, with effects similar to those on the momentum, here we want to verify if the same holds for compliant walls, that have not been studied so far. Thus, in this work we investigate the potential of compliant walls to promote heat transfer. To do this, we carry out direct numerical simulations of turbulent channel flows over fully-coupled compliant walls, in which we vary the wall elasticity and thermal diffusivity, comparing the results to those over a rigid wall. The manuscript is structured as follows: in \ref{sec:met} we introduce the mathematical formulation and numerical discretizations \cite{rosti_2026a}, then in \ref{sec:res} we discuss the results from the simulations, and finally in \ref{sec:con} we summarize and draw the main conclusions.

\section{Formulation and Numerical procedure} \label{sec:met}
The incompressible flow $\partial u^p_i/\partial x_i=0$ over compliant walls is governed by the Cauchy equation:
\begin{equation}
   \frac{\partial u_i^p}{\partial t} + \frac{\partial u^p_i u^p_j}{\partial x_j} = \frac{1}{\rho}\frac{\partial \sigma^p_{ij}}{\partial x_j}, \\
   \label{eq:momentum}
\end{equation}
where $u_i^p$ is the velocity field, $\sigma_{ij}^p$ the Cauchy stress tensor, and $\rho$ the density (assumed to be the same for both the solid and fluid phases). Here, we use the Einstein notation of repeated indices to indicate summation, and the suffix $p=f$ or $s$ to represent the fluid and solid phases. The Cauchy stress tensor is defined as
\begin{eqnarray}
   \sigma^f_{ij} &=& -p\delta_{ij} + 2\mu  \mathrm{D}_{ij}, \\[3pt]
   \sigma^s_{ij} &=& -p\delta_{ij} + 2\mu  \mathrm{D}_{ij} + G\mathrm{B}_{ij},
   \label{eq: stress}
\end{eqnarray}
where $p$ is the pressure field, $\mu$ is the dynamic viscosity (assumed to be the same in the two phases), $\mathrm{D}_{ij}$ is the strain rate tensor, and $\delta_{ij}$ is the Kronecker delta function. The last term in Eq. (\ref{eq: stress}) represents the hyperelastic contribution of the compliant wall, which we model as a neo-Hookean solid satisfying the incompressible Mooney-Rivlin law \cite{Bonet_Wood_2008}. $G$ indicates the transverse modulus of elasticity and $\mathrm{B}_{ij}$ is the left Cauchy–Green deformation tensor, found by enforcing that the upper convected derivative of $B_{ij}$ is null, i.e., $\overset{\triangledown}{\mathrm{B}}_{ij}=0$. Finally, we couple the previous set of equations with the following advection-diffusion equation for the temperature $T$,
\begin{equation}
    \frac{\partial T}{\partial t} + u_i \frac{\partial T}{\partial x_i} = \alpha \frac{\partial^2 T}{\partial x_j \partial x_j},
\end{equation}
where $\alpha$ is the thermal diffusivity.

The governing equations are solved in an Eulerian framework, as described in Refs.~\cite{Sugiyama2011,Rosti_Brandt_2017}. In particular, the multiphase problem is treated by a monolithic formulation based on the solid volume fraction $\phi_s$, an indicator function which is equal to $1$ in the solid, to $0$ in the fluid, and to $0.5$ at the interface. $\phi_s$ is found by solving an additional transport equation:
\begin{equation}
  \frac{\partial \phi_s}{\partial t} + \frac{\partial u_k \phi_s}{\partial x_k} = 0.
  \label{eq: transport}
\end{equation}
We advance in time the system of equations using a fractional step method based on the third-order Runge-Kutta scheme, except for the solid stress term in the Cauchy equation, which is advanced using the Crank-Nicolson scheme. All spatial derivatives are discretised using the second-order central finite-difference scheme, except for the advection terms in the transport equation for $B_{ij}$ and $\phi_s$, for which we use a WENO scheme \cite{Sugiyama2011}.

\begin{figure}[t]
     \centering
     \begin{subfigure}[c]{0.49\textwidth}
         \centering
         \includegraphics[width=0.9\textwidth]{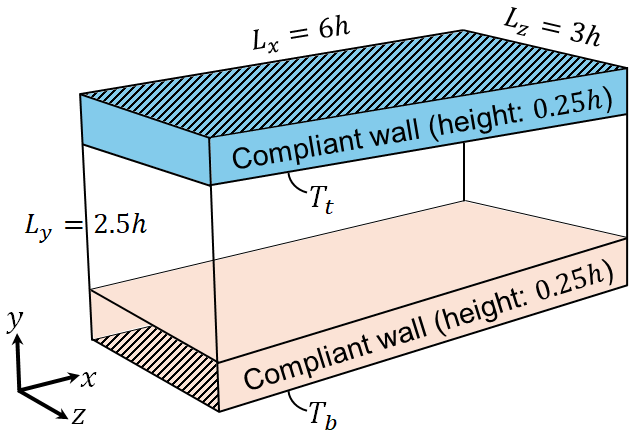}
         \caption{The computational domain}
         \end{subfigure}
     \hfill
     \begin{subfigure}[c]{0.49\textwidth}
     \begin{subfigure}[c]{0.48\textwidth}
         \centering
         \includegraphics[width=\textwidth]{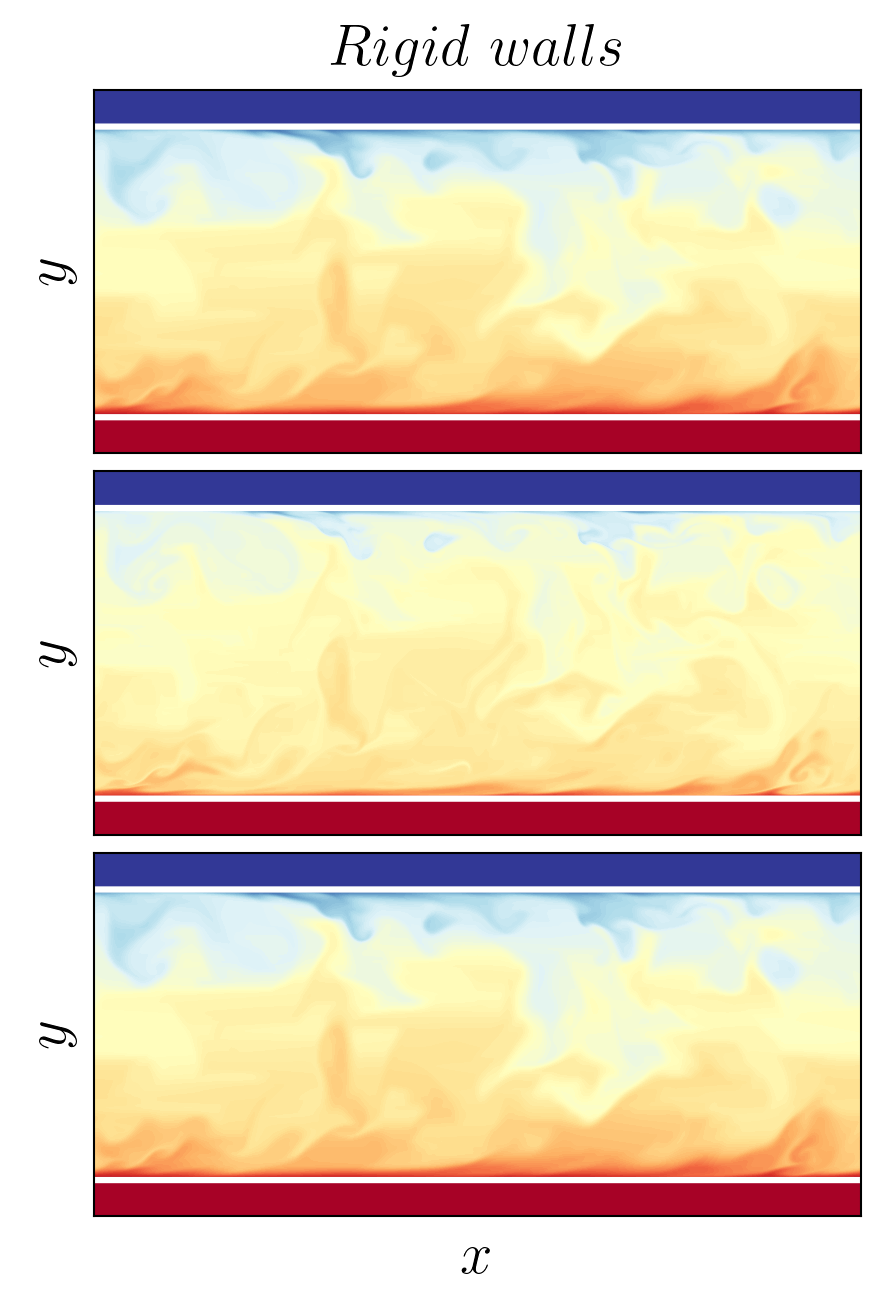}
         \end{subfigure}
     \hfill
     \begin{subfigure}[c]{0.48\textwidth}
         \centering
         \includegraphics[width=\textwidth]{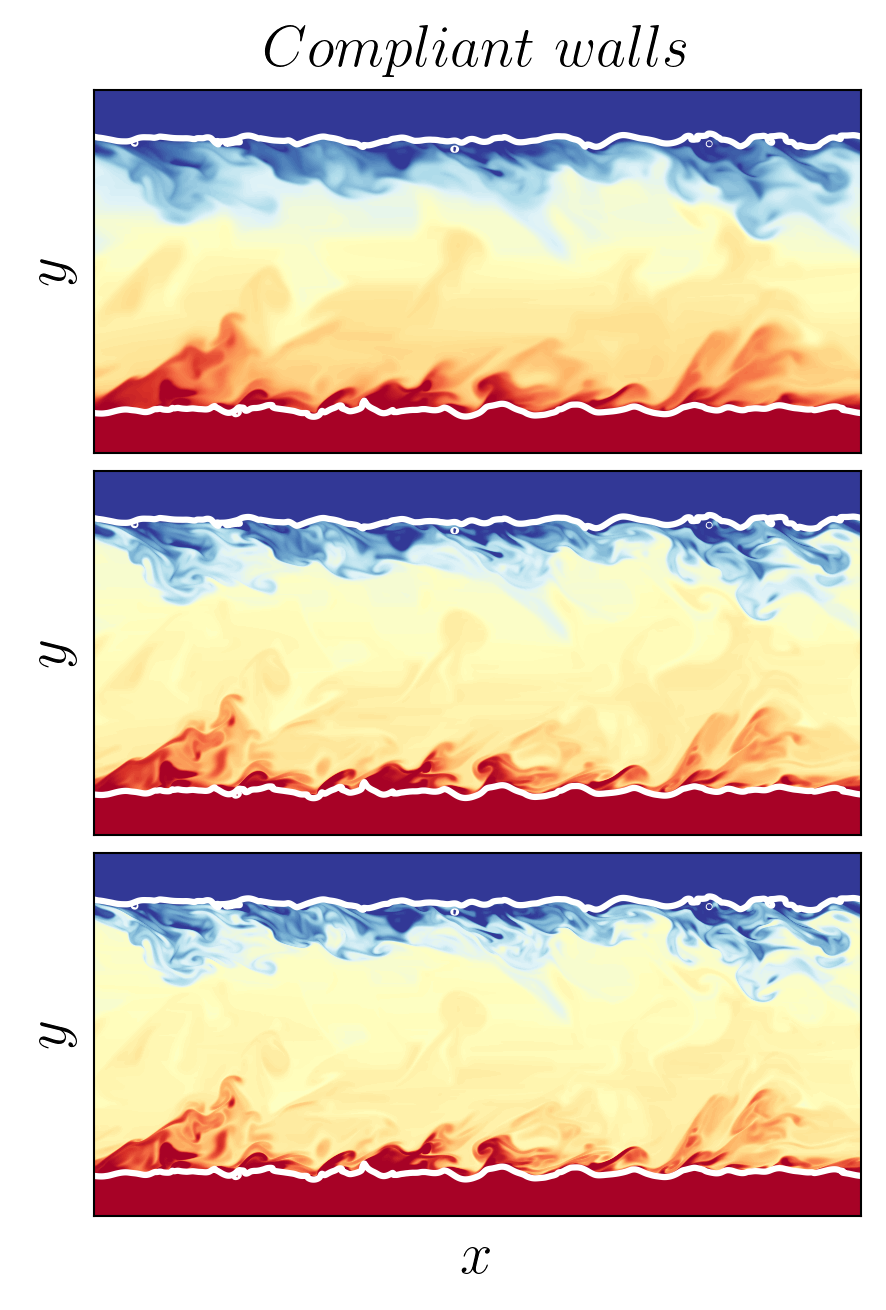}
         \end{subfigure}
     \caption{Instantaneous snapshot of temperature on $x$-$y$ planes}
     \end{subfigure}
     \caption{(a) Sketch of the computational configuration for the present study. The colored regions are the two compliant walls, bounded by two rigid-walls. (b) Visualizations of the instantaneous temperature on $x$-$y$ planes for varying Prandtl numbers: $Pr=1$ (top), $Pr=4$ (middle), and $Pr=7$ (bottom). The white solid lines represent the interface between fluid and compliant wall regions. The color-contours represent the temperature, $T$, with values ranging from $0$ (blue) to $1$ (red).}
    \label{fig: sketch}
\end{figure}

\subsection{Setup} \label{subsec:setup}
We study a fully developed turbulent channel flow of height $2h$, confined between two compliant solid walls of height $h_e=0.25h$ on the top and bottom, as shown in figure \ref{fig: sketch}. We indicate with $x$, $y$, and $z$ the streamwise, wall-normal, and spanwise directions. The compliant walls are initially flat, and are bounded by two impermeable rigid walls located at $y=-0.25h$ and $2.25h$. The size of the computational domain is $6h \times 2.5h \times 3h$, being discretised with a uniform cubic grid with $1920 \times 800 \times 960$ grid points, a domain size and spatial resolution comparable to the previous studies on the topic to ensure the reliability of the results \cite{Rosti_Brandt_2017, Ardekani_Rosti_Brandt_2019, rosti2020low, YOUSEFI2021Regimes, Koseki_Aswathy_Rosti_2025}. We impose periodic boundary conditions in the streamwise and spanwise directions, and the no-slip condition in the wall-normal direction on the rigid walls. Continuity of the velocity and normal traction is enforced across the interface between the fluid and elastic solid. A constant flow rate is enforced in the streamwise direction by updating the pressure gradient at every time step to ensure a fixed bulk velocity $U_b$. Finally, we apply a constant temperature difference $\Delta T$ as the thermal boundary condition on the two compliant walls. Note that, we formally solve the equation of the temperature in the whole domain, i.e., in both the solid and fluid phases, using the mixture temperature $T=T_s\phi_s + T_f \left( 1- \phi_s \right)$, in analogy to what is done for the velocity and pressure fields. To enforce the desired temperature in each of the compliant walls, we add a forcing term $\phi_s\left( T-T_w \right)/\Delta t $, ($\Delta t$ representing the time increment) to the temperature equation, similarly to what is done to enforce the constant bulk velocity in the velocity field (where the forcing term $\left( 1- \phi_s \right) \left( \int v~d\mathcal{V}-U_b \right)/\Delta t$ is added to the streamwise momentum equation). With this procedure, the temperature inside the solid wall is not perfectly stationary, but the variations are sufficiently small.

With the above setup and assumptions, the flow is governed by three non-dimensional parameters: \textit{i)} the Reynolds number $Re_b=U_b h/\nu$ that we fix at $3500$; \textit{ii)} the non-dimensional transverse elastic modulus $G/(\rho U_b^2)$, for which three values are simulated, $0.25$ (very deformable), $0.5$ (almost rigid), and $\infty$ (rigid); \textit{iii)} the Prandtl number $Pr=\nu/\alpha$ (with $\nu$ being the kinematic viscosity $\nu=\mu/\rho$), which is set equal to $Pr=1$, $4$, and $7$. Note that, we select the values of the wall elasticity to range conditions in which the timescale of the elastic waves $h_e\sqrt{\rho/G}$ are larger, comparable, and smaller than the timescale of the bulk flow $h/U_b$.

\section{Results} \label{sec:res}
We simulate the turbulent heat transfer over compliant walls to understand the effects of the wall motion on the temperature field. In section \ref{sec:a}, we start comparing the results for different Prandtl numbers for a rigid and compliant wall -- the most elastic considered here --, later move to assess the influence of different wall elasticity in \ref{sec:b}, and finally we investigate the similarity between heat and momentum transfer in \ref{sec:c}.

\subsection{Rigid vs compliant walls} \label{sec:a}

\begin{figure}[tb]
     \centering
     \begin{subfigure}[c]{0.49\textwidth}
         \centering
         \includegraphics[width=\textwidth]{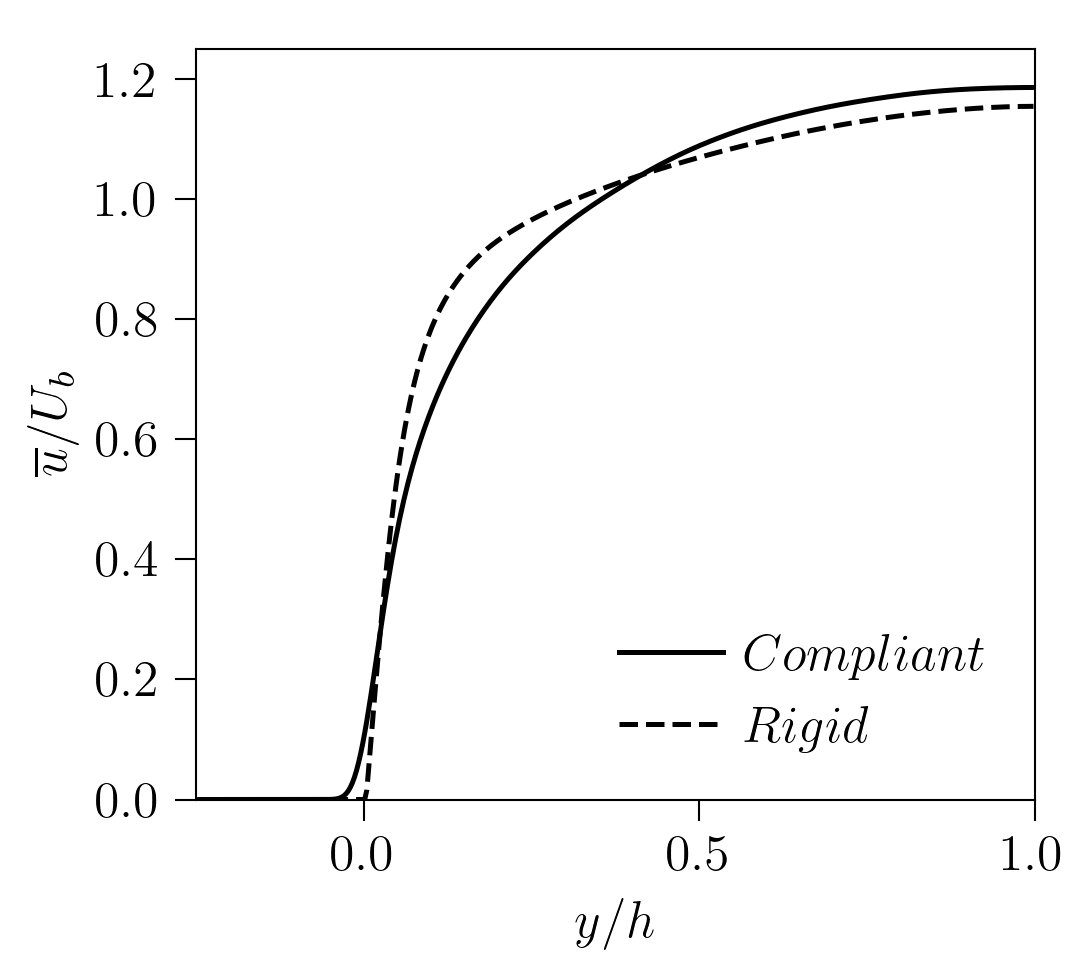}
         \end{subfigure}
     \hfill
     \begin{subfigure}[c]{0.49\textwidth}
         \centering
         \includegraphics[width=\textwidth]{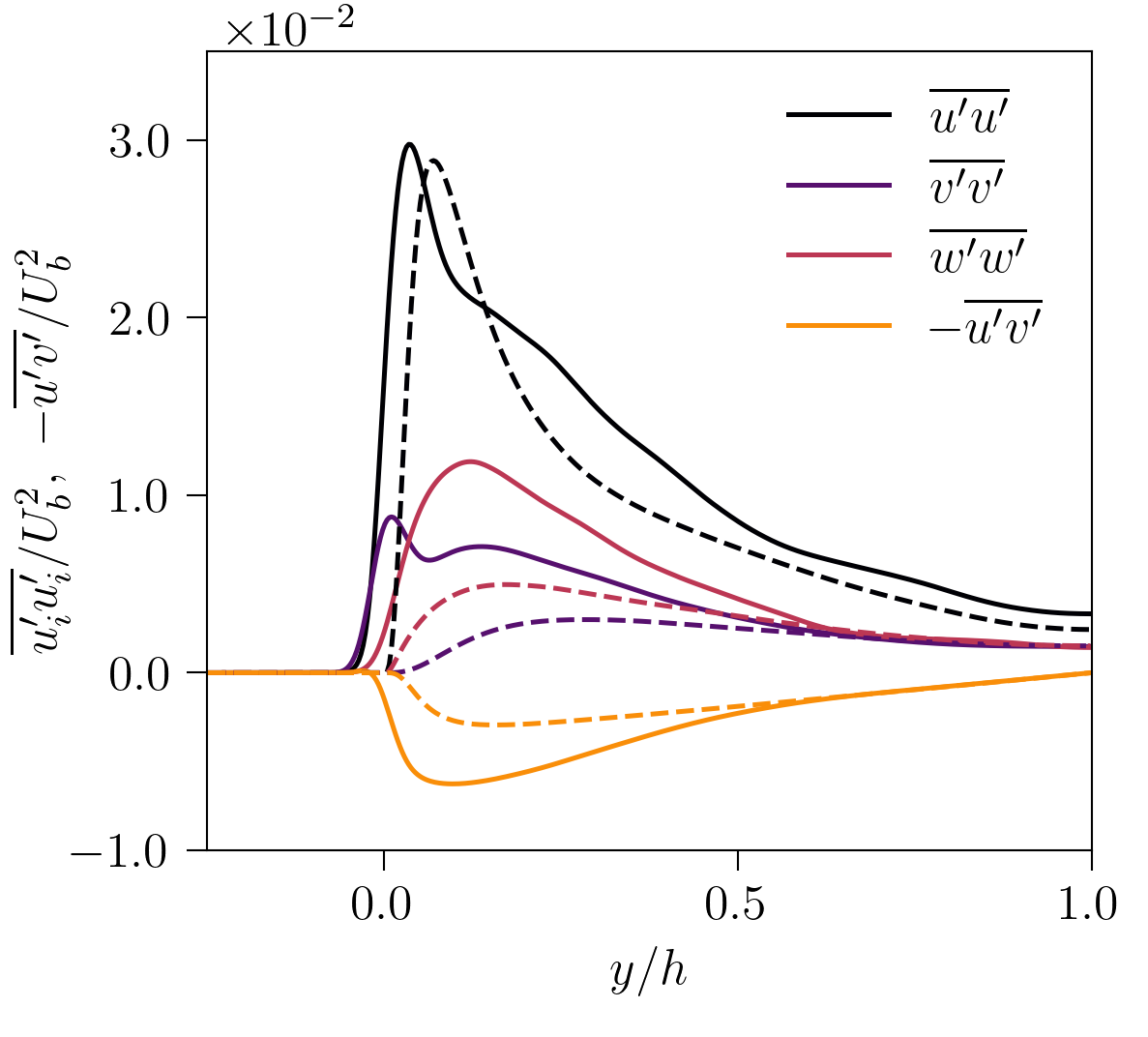}
         \end{subfigure}
        \caption{The wall-normal profiles of the flow statistics: (left) mean velocity profiles and (right) Reynolds shear and normal stress components of the fluid. The line style differentiate the cases with different walls: solid line for the compliant wall with $G=0.25\rho U_b^2$ , and dashed line for the rigid wall. The line color in the right panel indicates the different components of the Reynolds stress tensor: $\overline{u^\prime u^\prime}$ (black), $\overline{v^\prime v^\prime}$ (violet), $\overline{w^\prime w^\prime}$ (magenta), and $\overline{u^\prime v^\prime}$ (orange).} 
        \label{fig: fluid statistics}
\end{figure}

The effect of a compliant wall on a turbulent channel flow has been recently investigated \cite{Rosti_Brandt_2017,Ardekani_Rosti_Brandt_2019,Esteghamatian2022spatiotemporal,Koseki_Aswathy_Rosti_2025}, thus we only briefly report here the main turbulent flow statistics for the sake of completeness. Note that, in the following $\overline{(\cdot)}$ indicates the average in time and along the streamwise and spanwise directions, except otherwise noted, while $(\cdot)^\prime$ represents the fluctuations.

Figure \ref{fig: fluid statistics} shows the profiles of the (left) streamwise mean velocity and (right) normal and shear components of the Reynolds stress tensor. Due to the compliant wall, both the mean velocity and the Reynolds stress tensor are strongly altered with respect to the case of a rigid wall, with the flow modification being in close agreement with those previously reported \cite{Rosti_Brandt_2017, Ardekani_Rosti_Brandt_2019, Esteghamatian2022spatiotemporal, Koseki_Aswathy_Rosti_2025}. In particular, the mean streamwise velocity exhibits an increase of the maximum velocity at the centerline, and a reduction of the shear at the wall, leading to a reduction of the viscous shear stress at the wall. However, total friction increases significantly in the elastic wall case, as proved by the corresponding friction Reynolds numbers equal to $323$, $264$, and $219$ for $G/(\rho U_b^2)=0.25,\ 0.5,$ and $\infty$, respectively. This increase is due to the overall increase of the Reynolds shear stress component shown in figure \ref{fig: fluid statistics}(right), which is not null at the wall in the elastic case due to the wall oscillations. In general, the compliant wall enhances all the components of the Reynolds stress tensor throughout the channel, ultimately vanishing only inside the elastic layer. This is especially true for the wall-normal and spanwise components, while the change of streamwise component remains rather limited. Note also the presence of secondary peaks in the wall-normal component located close to the mean interface location, which is due to the oscillation of the elastic wall \cite{Rosti_Brandt_2017}. 
As discussed in Ref.~\cite{Koseki_Aswathy_Rosti_2025}, the flow modifications are not simply due to the wall shape acting as roughness, but mostly arise from the continuous wall movement.

\begin{figure}[tb]
     \centering
     \begin{subfigure}[c]{0.49\textwidth}
         \centering
         \includegraphics[width=\textwidth]{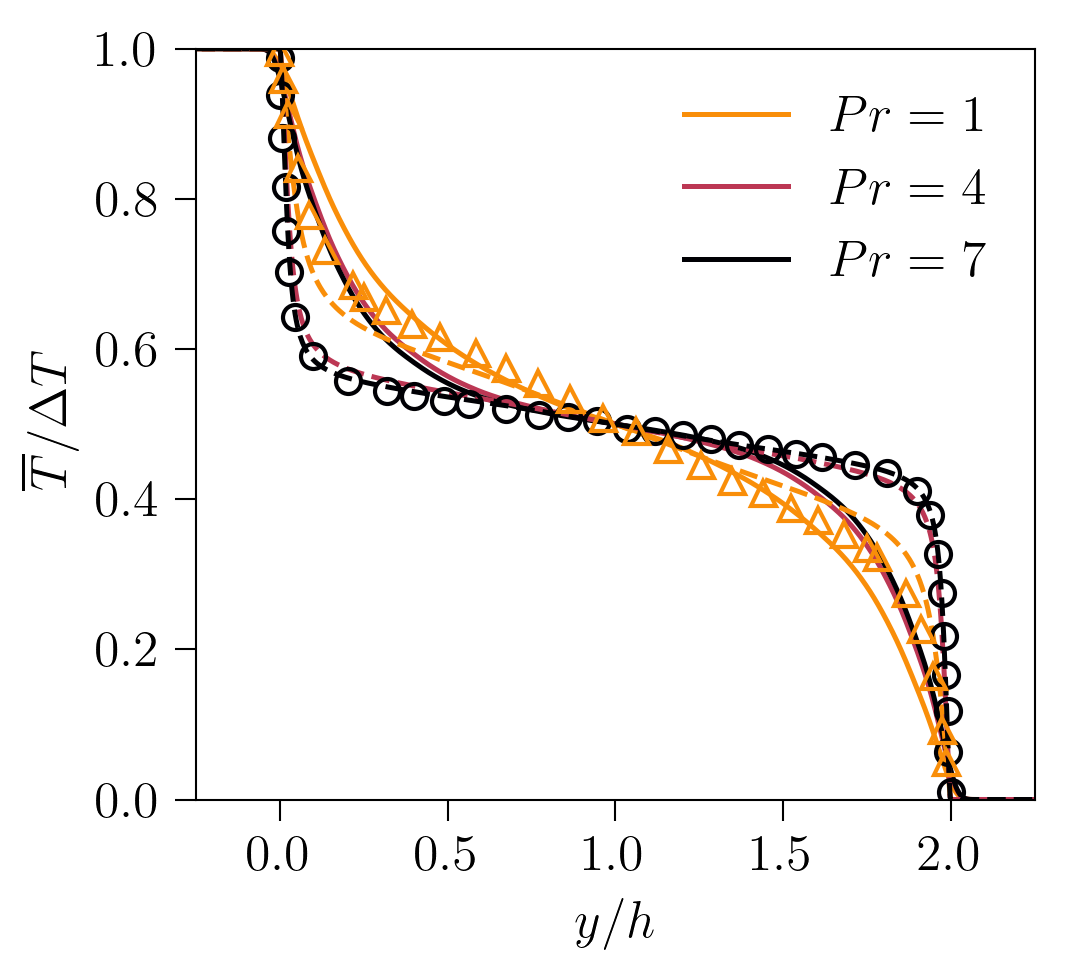}
         \end{subfigure}
     \hfill
     \begin{subfigure}[c]{0.49\textwidth}
         \centering
         \includegraphics[width=\textwidth]{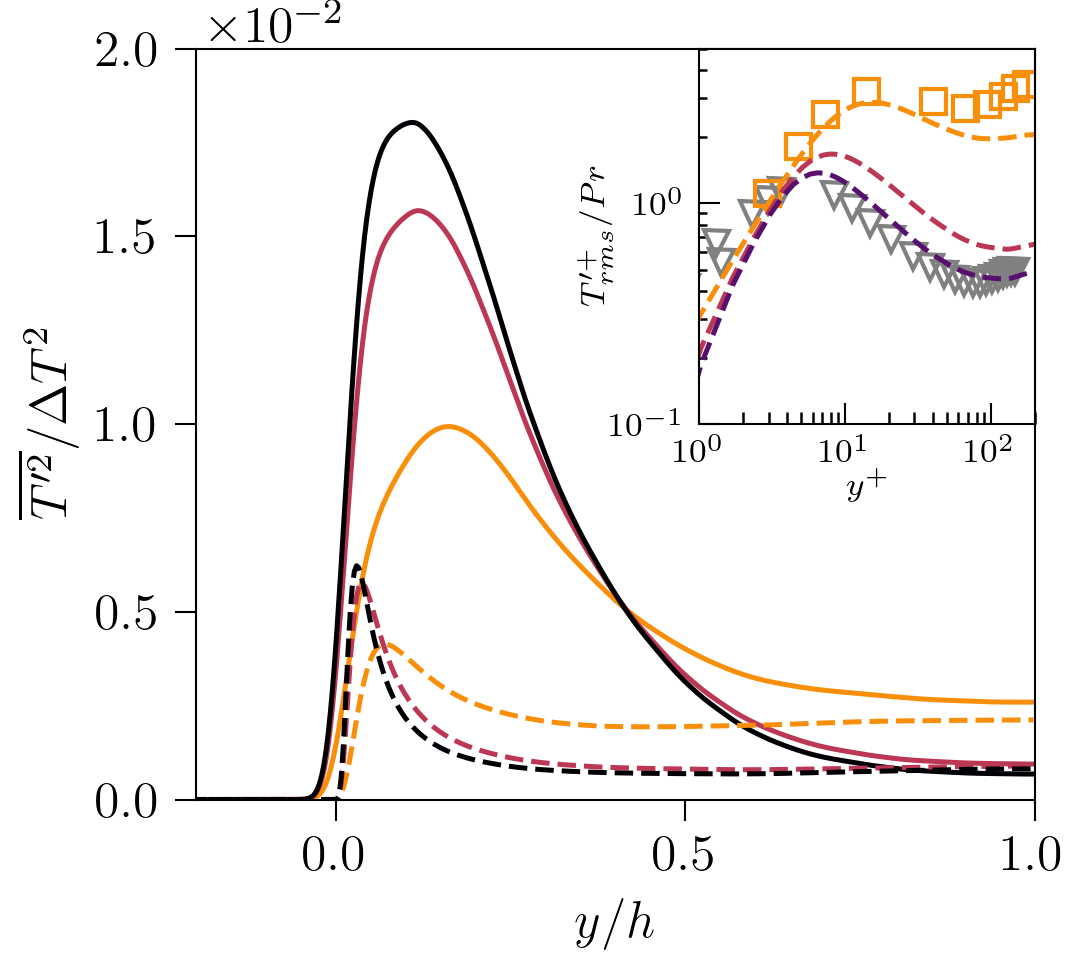}
         \end{subfigure}
        \caption{The wall-normal profiles of the thermal statistics: (left) mean temperature profiles and (right) variance of the temperature fluctuations. The line style differentiate the cases with different walls: solid line for the compliant wall with $G=0.25\rho U_b^2$, and dashed line for the rigid wall. The line color indicates the different Prandtl number: $Pr=1$ (orange), $Pr=4$ (magenta), and $Pr=7$ (black). Symbols represent the reference results from the literature: (left) $\triangle$: $Pr=1$ at $Re_\tau=150$ \cite{NA1999use}, and $\bigcirc$: $Pr=7$ at $Re_b=2800$ \cite{YOUSEFI2021Regimes}; (right) $\Box$: $Pr=1$ at $Re_\tau=180$ \cite{ZHOU2025dns}, and $\bigtriangledown$: $Pr=10$ at $Re_\tau=150$ \cite{NA1999use}.}
        \label{fig: temp statistics}
\end{figure}

Similarly to the flow field, compliant walls strongly affect the thermal field. Figure \ref{fig: temp statistics} (left) shows the mean temperature profiles for the compliant and rigid walls at different Prandtl numbers. For both cases, as the Prandtl number increases, the gradient of the temperature at the wall is enhanced, consistently with the results from the literature for rigid-walls reported in the figure with symbols \cite{PAPAVASSILIOU1997transport, NA1999use}. When the wall is compliant however, the gradient at the wall is reduced compared to the rigid-wall case at the same $Pr$, suggesting a reduction in the diffusive thermal transfer. The decrease in the temperature gradient near the wall can be linked to the diminished velocity variation in the same region, previously shown in figure \ref{fig: fluid statistics}(left), and is qualitative similar to what observed over rough and porous walls \cite{nishiyama2020direct}. Figure \ref{fig: temp statistics}(right) reports the variance of the temperature field. In the rigid-wall case, the temperature fluctuations have a peak close to the wall, followed by an almost constant value in the bulk with a small secondary peak at the center of the channel. Increasing the Prandtl number does not alter the qualitative behaviour \cite{NA1999use}, with only an enhancement of the near wall peak which displaces closer to the wall. Note also that the effect of the Prandtl number seems to saturate beyond $Pr=4$. The observed behaviour for the rigid wall is in good agreement with the results from the literature \cite{NA1999use, ZHOU2025dns}, as shown in the inset of figure \ref{fig: temp statistics}(right). In the compliant-wall case, we observe a substantial enhancement of the peak, which broadens and becomes more intense; note that enhanced temperature fluctuations occupy more than half of the channel height. The Prandtl number alters the picture similarly to what done with the rigid wall, with a stronger peak located closer to the wall as $Pr$ increases. While strong temperature fluctuations are located within a region of large temperature gradients in the rigid case, the intense temperature fluctuations in the compliant case are concentrated further away from the wall, suggesting that the heat transfer may be driven by a different mechanism when the wall is compliant.
 
\begin{figure}[tb]
     \centering
     \begin{subfigure}[c]{0.49\textwidth}
         \centering
         \includegraphics[width=\textwidth]{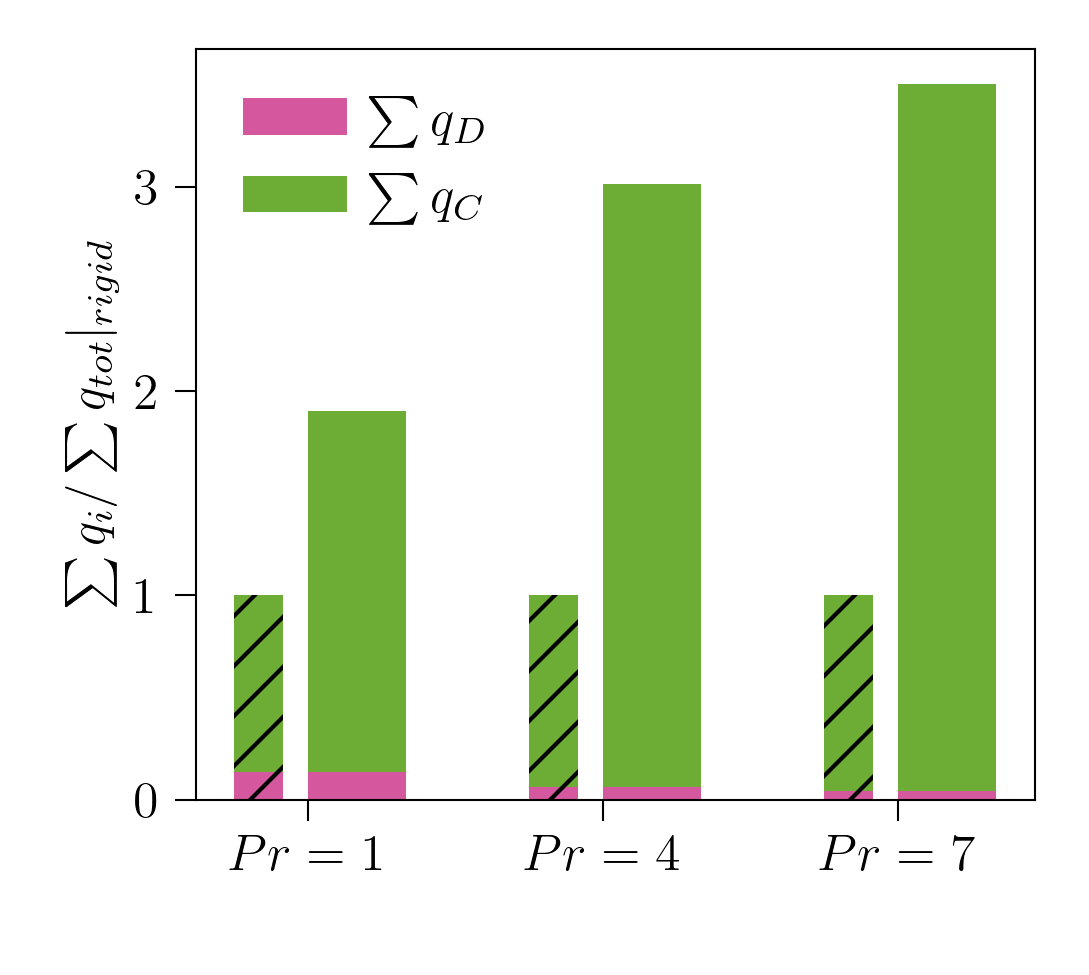}
         \end{subfigure}
     \hfill
     \begin{subfigure}[c]{0.49\textwidth}
         \centering
         \includegraphics[width=\textwidth]{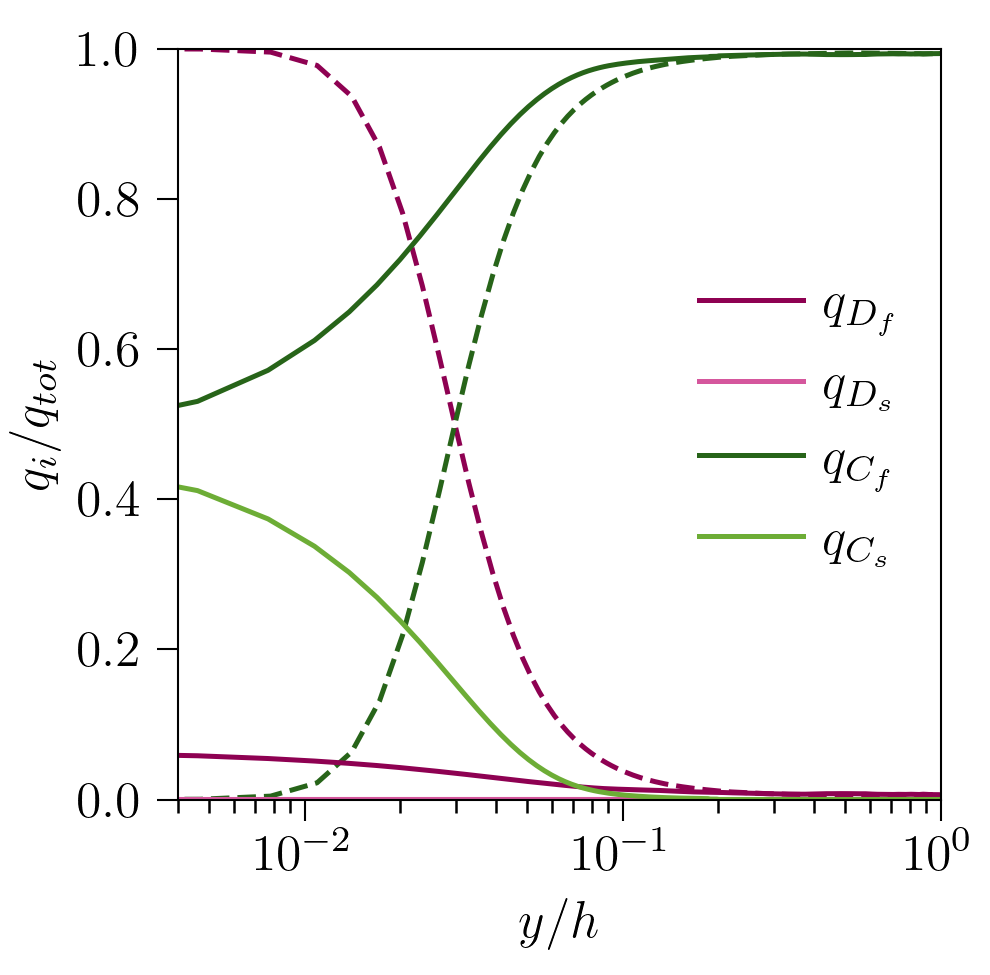}
         \end{subfigure}
        \caption{(left) The integral contribution of the total diffusive $q_{D}$ (magenta) and convective $q_{C}$ (green) heat flux components introduced in Eq.~(\ref{eq: heat flux}) for different Prandtl numbers. The rigid case is shown with thinner bars and a black hatched fill. (right) The wall-normal profiles of all the heat flux contributions for $Pr=7$, with each term normalized by the local total heat flux. The line color represents the different terms in Eqs. (\ref{eq: dtdy term}) and (\ref{eq: vt term}): $q_{D_f}$ (dark magenta), $q_{D_s}$ (light magenta), $q_{C_f}$ (dark green), and $q_{C_s}$ (light green). The rigid case is shown with dashed lines for comparison.}
        \label{fig: heat flux}
\end{figure}

To understand and quantify how the above observations affect the heat transfer, we look at the heat transfer budget. The total heat flux $q_{tot}$ is composed of two contributions: the viscous diffusion $q_D$ and the convection heat flux $q_C$ resulting from the correlation of velocity and temperature fluctuations. In our case, each contribution can also be divided into the fluid and solid phases, distinguished with the subscripts $_f$ for the fluid and $_s$ for the solid. The total heat flux is thus determined by:

\begin{equation}
    q_{tot} = q_{D} + q_{C},
    \label{eq: heat flux}
\end{equation}
with each term expressed as:
\begin{equation}
    q_D = q_{D_f} + q_{D_s} = \overline{(1-\phi^s)  \alpha \frac{dT}{\mathop{dy}}} + \overline{\phi^s  \alpha \frac{dT}{\mathop{dy}}},
    \label{eq: dtdy term}
\end{equation}
\begin{equation}
    q_C = q_{C_f} + q_{C_s} = \overline{(1-\phi^s)T^\prime v^\prime} + \overline{\phi^s T^\prime v^\prime}.
    \label{eq: vt term}
\end{equation}
Note that, the temperature is kept fixed in the solid phase through a forcing term, as explained in \ref{subsec:setup}, thus the fluxes basically vanishes inside the bulk of the solid phase. At the interface where the temperature is a mixture, the solid-diffusive term still remains essentially null, due to the small variation of the temperature mean field, while the solid-convection term cannot be ignored, due the compliant-wall movement and the temperature fluctuations. We can thus consider this latter term as a proxy measure of the wall movement.

Figure \ref{fig: heat flux}(left) shows the integral contribution of each term across the wall-normal direction, normalized by the integral heat flux of the corresponding case with rigid walls. The figure clearly shows that the total heat flux increases significantly in the presence of compliant walls, with the major difference coming from the convection term which doubles for $Pr=1$ and triples for $Pr=7$, while the integral contribution of the viscous diffusion does not demonstrate significant differences compared to the rigid case, being only slightly reduced and decreasing with $Pr$ for both the rigid and compliant cases. To understand the origin of these modifications, figure \ref{fig: heat flux}(right) shows the vertical profiles of the different heat transfer mechanisms across the channel, further separated into the contribution of the fluid and solid phases. In the rigid-wall case, heat transfer is driven by both diffusion and turbulent convection heat flux: diffusion is dominant in the near-wall region, whereas convection overtakes and dominates in the bulk of the channel away from the wall (see e.g., Fig.7 (a) in Ref. \cite{Ardekani_Rosti_Brandt_2019}). On the other hand, in the compliant case, the contributions from both diffusion (of the fluid) and convection of the solid are dominant in the near-wall region, with the solid convection term being much larger than the viscous diffusion. Note that, the viscous diffusion in the solid is null. Away from the wall, the fluid convection rises, replacing the reduction of the solid-phase contributions. These results prove that the compliant walls significantly affect the whole near-wall heat transfer process, by replacing the diffusive contribution at the wall with the combination of fluid and solid convection, that arise from the movement of the wall pumping momentum in the bulk of the channel, transporting the temperature at the same time.

\begin{figure}[tb]
     \centering
     \begin{subfigure}[c]{\textwidth}
         \centering
         \includegraphics[width=\textwidth]{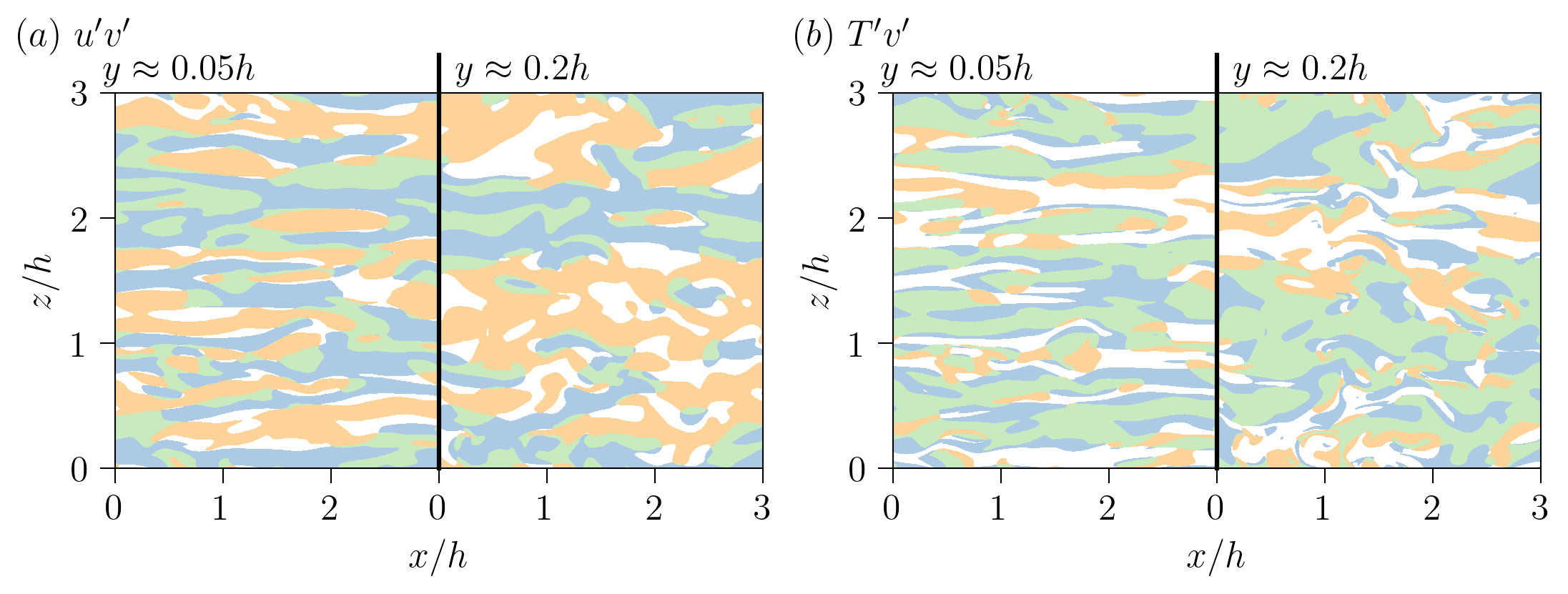}
         \end{subfigure} \\
     \begin{subfigure}[c]{\textwidth}
         \centering
         \includegraphics[width=\textwidth]{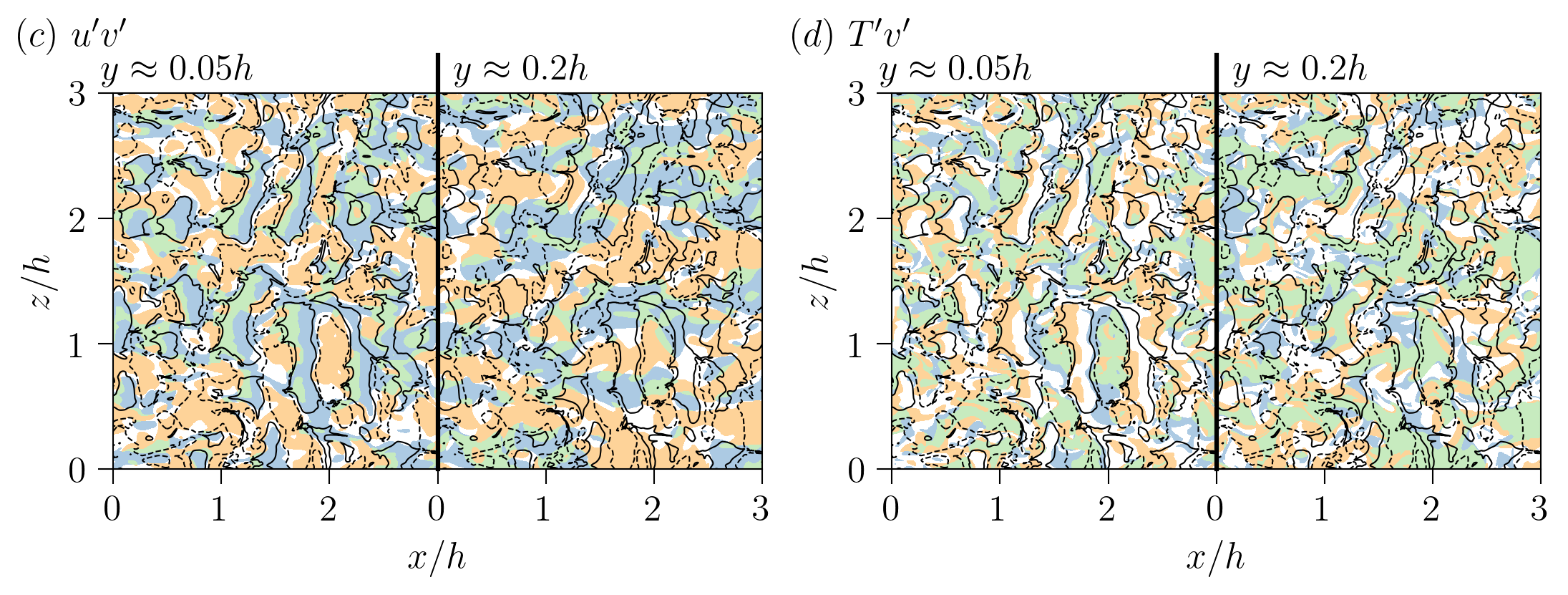}
         \end{subfigure} \\
     \begin{subfigure}[c]{\textwidth}
         \centering
         \includegraphics[width=\textwidth]{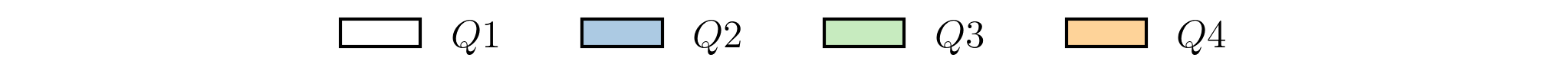}
         \end{subfigure}
     \caption{Instantaneous visualization on $x-z$ planes at $y\approx0.05h$ and $0.2h$ of the quadrant events for (a, c) $u^\prime v^\prime$ and (b, d) $T^\prime v^\prime$ at $Pr=7$. The top and bottom panels are for the rigid (a, b) and compliant (c, d) cases. Thin black lines represent the instantaneous iso-contour-lines of wall deformation $\delta/h$: $-0.15$ (dashed) and $0.15$ (solid).}
     \label{fig: instantaneous 2d events}
\end{figure}

\begin{figure}[tb]
     \centering
     \begin{subfigure}[c]{0.49\textwidth}
         \centering
         \includegraphics[width=0.7\textwidth]{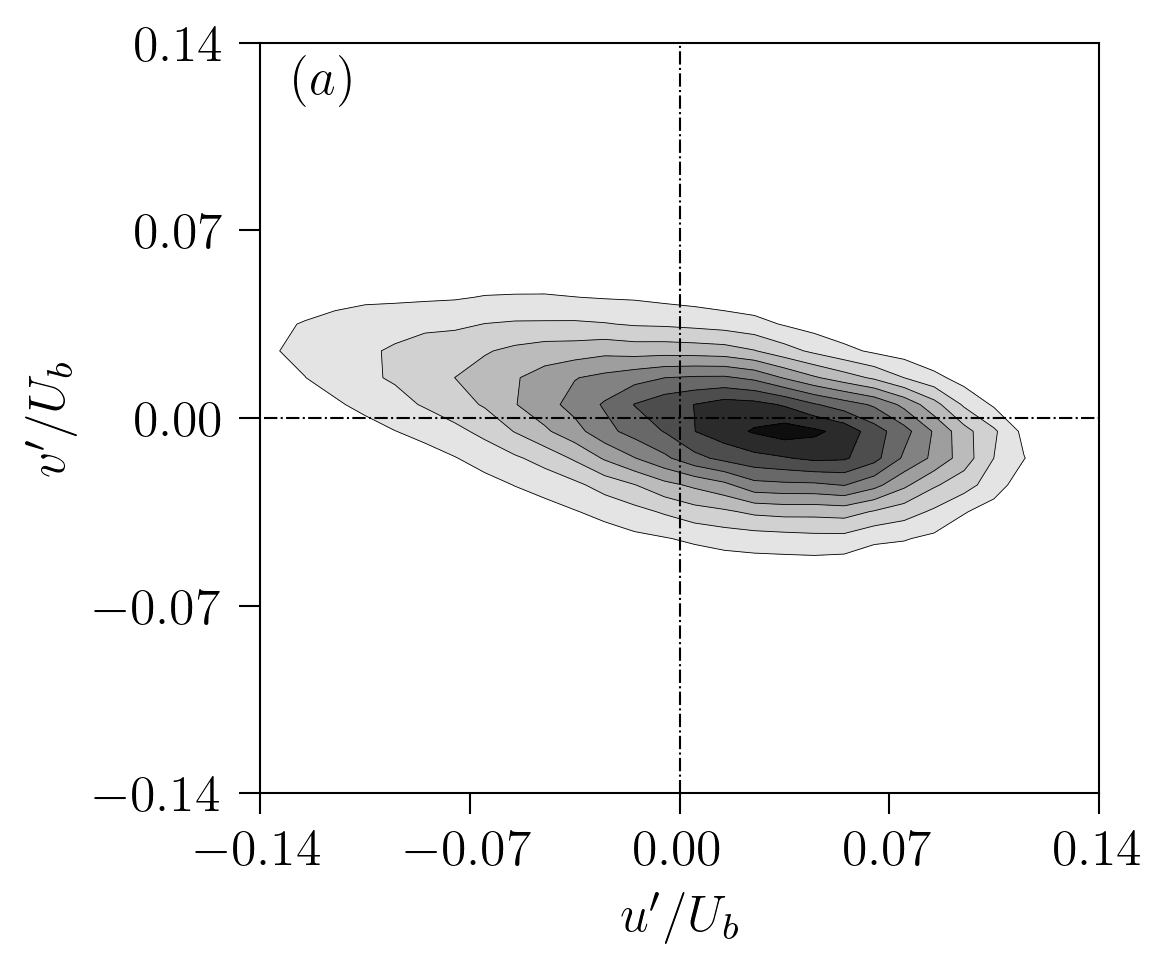}
         \end{subfigure}
     \hfill
     \begin{subfigure}[c]{0.49\textwidth}
         \centering
         \includegraphics[width=0.7\textwidth]{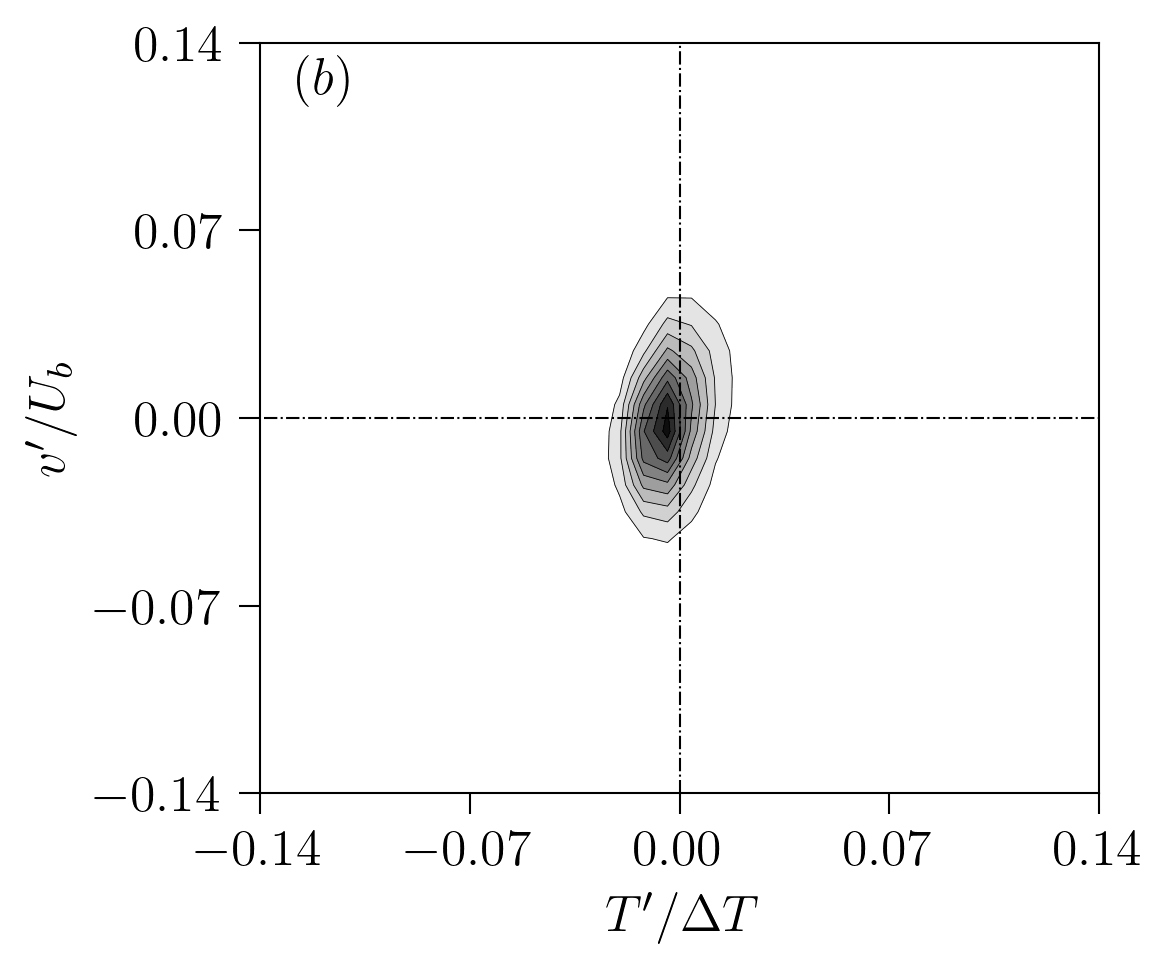}
         \end{subfigure} \\
     \begin{subfigure}[c]{0.49\textwidth}
         \centering
         \includegraphics[width=0.7\textwidth]{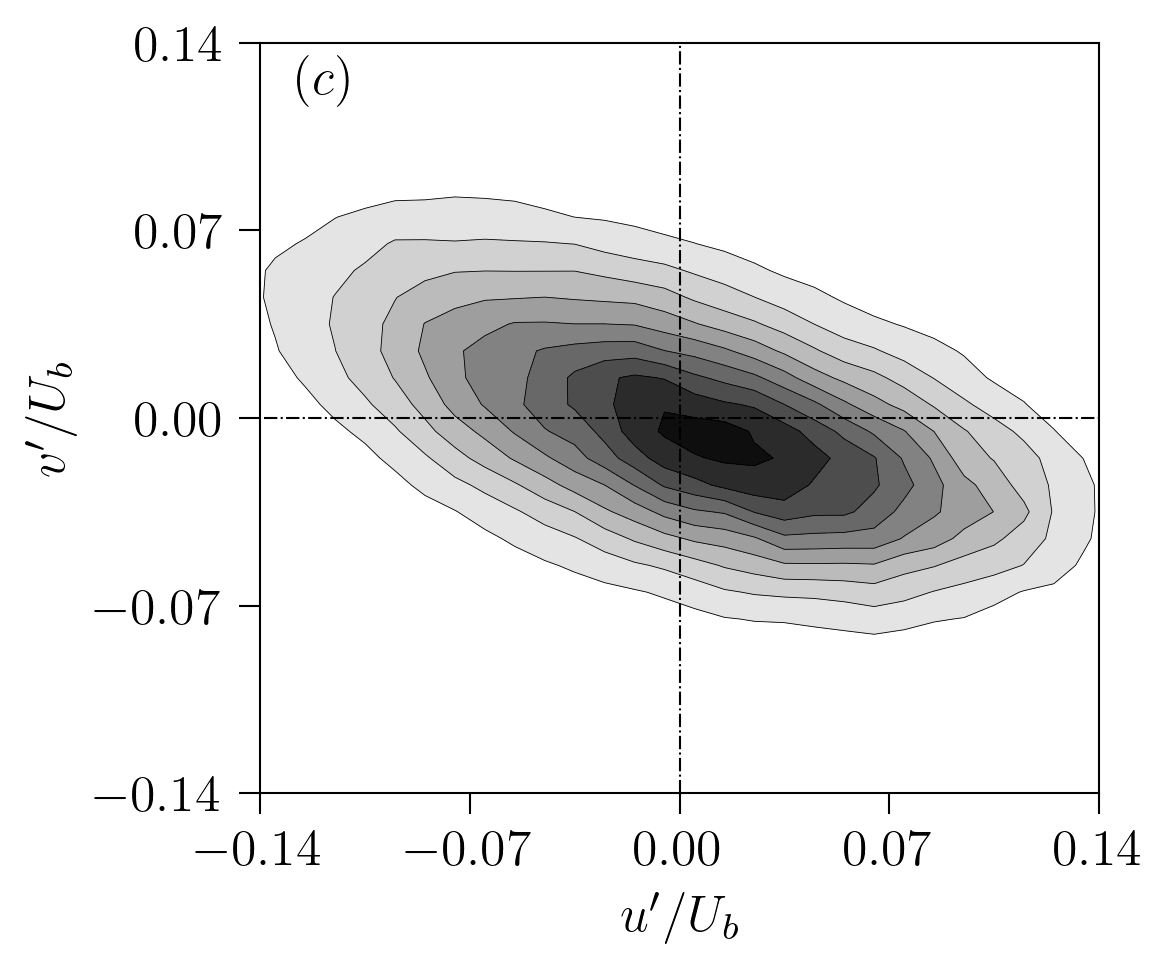}
         \end{subfigure}
     \hfill
     \begin{subfigure}[c]{0.49\textwidth}
         \centering
         \includegraphics[width=0.7\textwidth]{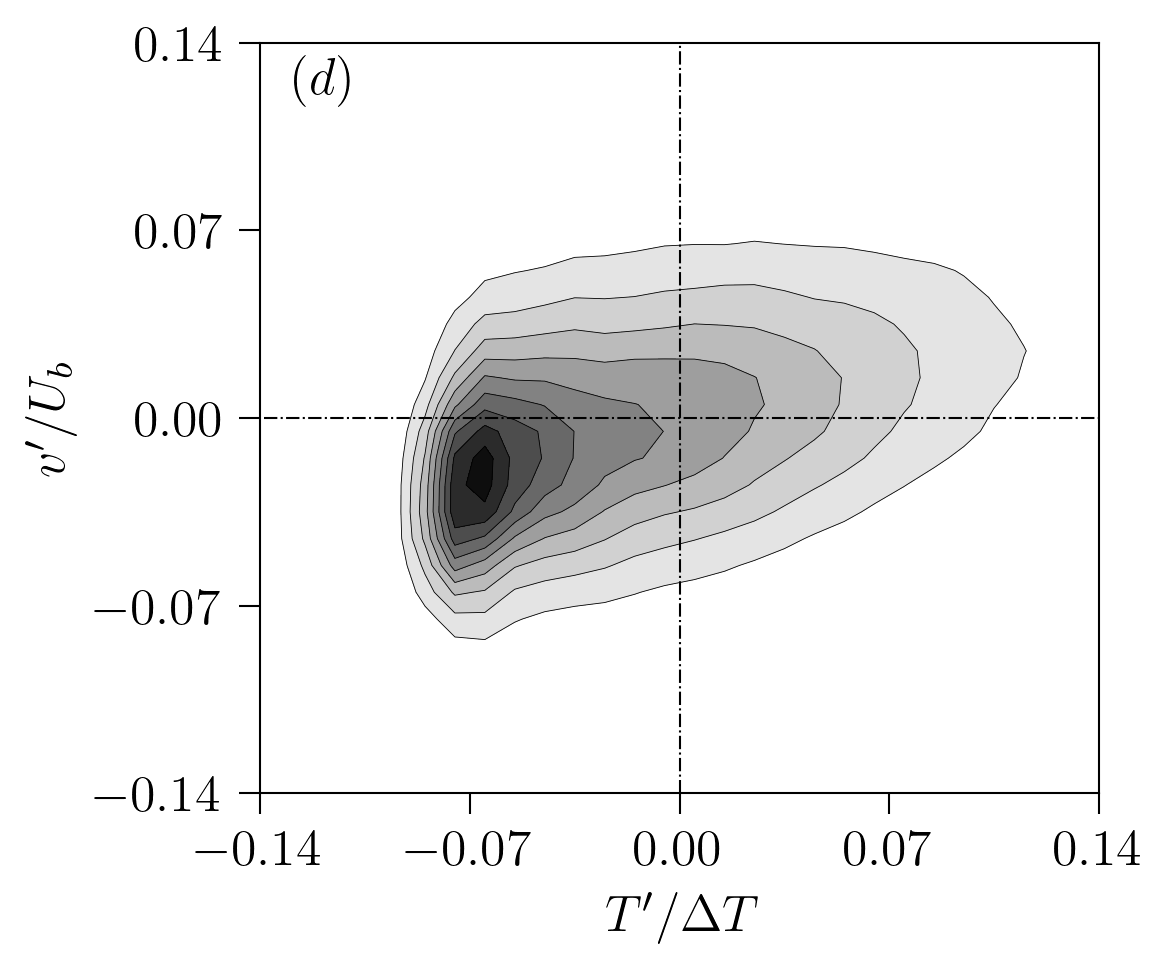}
         \end{subfigure}
     \caption{Joint probability density function of the (a, c) velocity fluctuations $u^\prime$ and $v^\prime$ and of the (b, d) temperature $T^\prime$ and wall-normal velocity $v^\prime$ fluctuations at $Pr=7$, extracted on a $x-y$ plane at $y\approx0.2h$. The top and bottom panels are for the rigid (a, b) and compliant (c, d) cases. The contour levels range from $0.15$ to $0.95$ with 0.1 increments.} 
     \label{fig: joint pdf}
\end{figure}

We next investigate the correlation between the velocity and temperature fluctuations. We do this by using the quadrant analysis and decomposing the $u^\prime v^\prime$ ($T^\prime v^\prime$) correlation into four quadrants \cite{katul1997ejection}: Q1, positive $u^\prime$ ($T^\prime$) and $v^\prime$; Q2, negative $u^\prime$ ($T^\prime$) and positive $v^\prime$; Q3, negative $u^\prime$ ($T^\prime$) and $v^\prime$; Q4, positive $u^\prime$ ($T^\prime$) and negative $v^\prime$. Note that a positive momentum convective flux corresponds to Q2 and Q4, while a positive heat flux corresponds to Q1 and Q3. 

Figure \ref{fig: instantaneous 2d events} shows the instantaneous distribution of the events of $u^\prime v^\prime$ and $T^\prime v^\prime$ on $x-z$ planes at different wall-normal heights ($y/h \approx 0.05$ and $0.2$). For the momentum flux, the event distributions highlight the coherent turbulent structures populating the velocity field, which are organized along the streamwise direction above the rigid-wall, whereas they tend to be fragmented and shorter in length above the compliant-wall case, with an enhanced spanwise coherency, as reported before \cite{Rosti_Brandt_2017, Koseki_Aswathy_Rosti_2025}. The above trend is clear especially in the near wall region (at $y/h \approx 0.05$), and the resulting structures well correlate with the wall deformation: downward motions of the compliant wall (dashed line) correspond to sweep events (orange-colored region), while upward motions (solid line) correspond to ejections (blue-colored region), as also reported by Ref. \cite{Ardekani_Rosti_Brandt_2019}. Farther from the wall ($y/h \approx 0.2$), the structures merge and grow in size in both cases. These modifications for the compliant wall, thus, can be linked to the wall oscillations generating intense wall-normal fluctuations that break down the streamwise coherence of the flow. For both walls, the majority of the events are in Q2 and Q4 (positive convective flux of momentum).

Similar observations hold for the thermal flux, with $Q1$ and $Q3$ being the most probable events (positive convective flux of heat), and with the structures becoming shorter and less elongated in the streamwise direction over the compliant wall. Furthermore, in the compliant-wall case, we observe that downward wall deformations drag low-temperature fluid close to the wall (dashed line and green-colored region), while upward wall deformations bring the heated fluid back into the bulk region (solid line and white-colored region). We can observe very strong similarity in the shape and patterns of the events of momentum and heat convective fluxes; this is easily ascribed to the fact that the temperature field is carried by the turbulent flow, and indeed it has been previously reported for rigid \cite{kim1989transport, Kasagi1992direct, Kong2000direct}, permeable \cite{nishiyama2020direct}, and rough  \cite{Leonardi_Orlandi_Djenidi_Antonia_2015} walls. The same also holds here for the flow over compliant walls, where we have shown that turbulent convection is the dominant process.

\begin{figure}[h]
     \centering
     \begin{subfigure}[c]{0.44\textwidth}
         \centering
         \includegraphics[width=\textwidth]{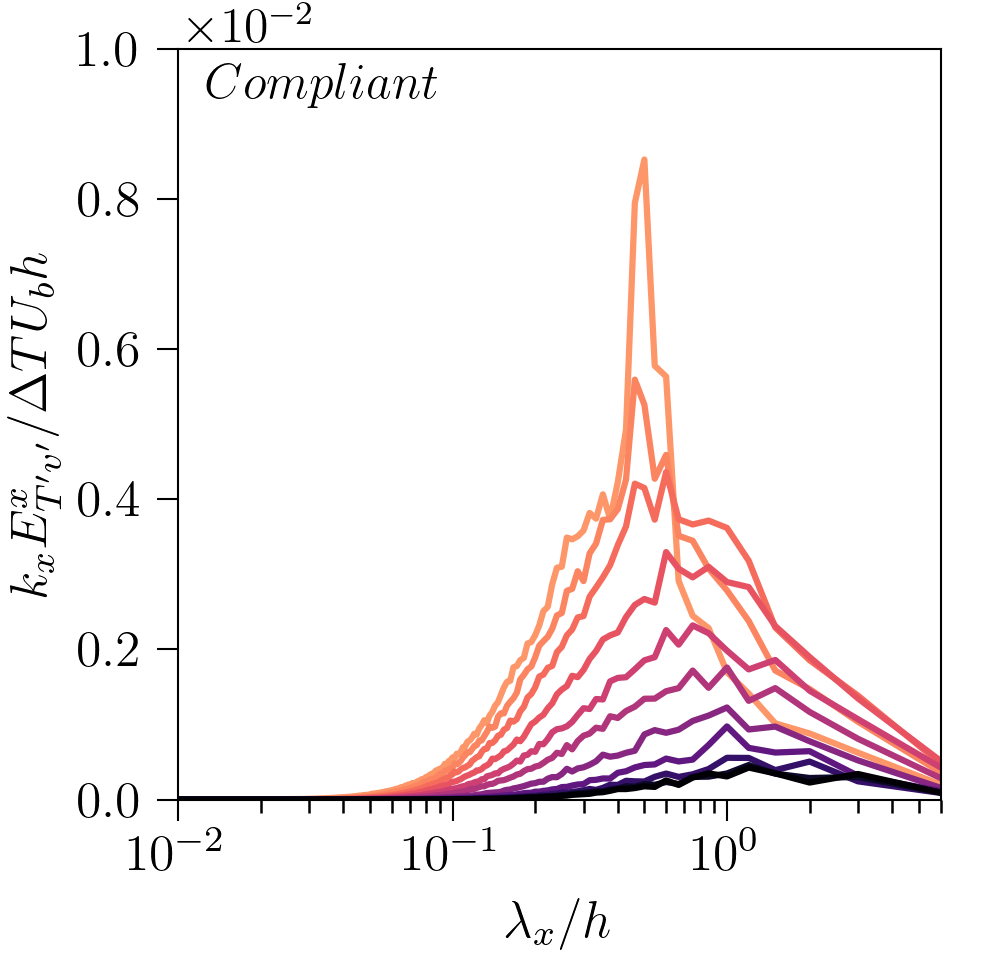}
         \end{subfigure}
     \begin{subfigure}[c]{0.44\textwidth}
         \centering
         \includegraphics[width=\textwidth]{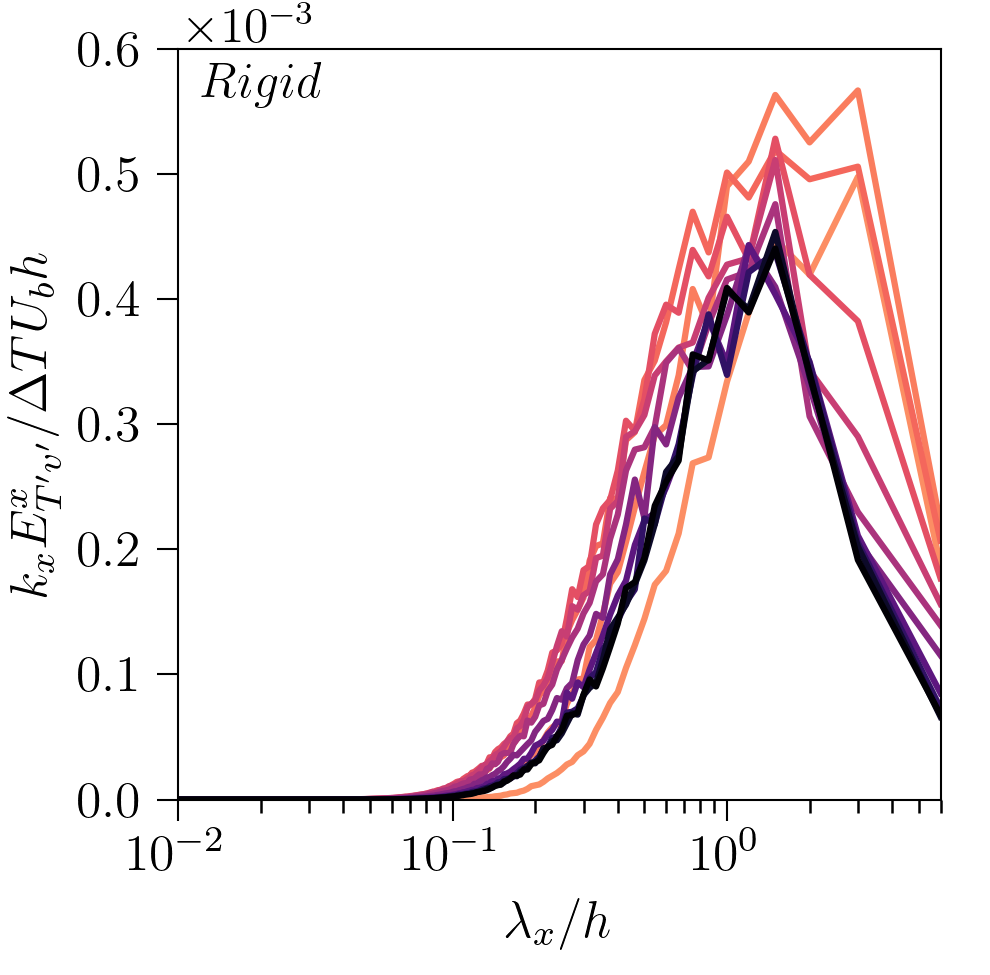}
         \end{subfigure}
     \begin{subfigure}[c]{0.1\textwidth}
         \centering
         \includegraphics[width=\textwidth]{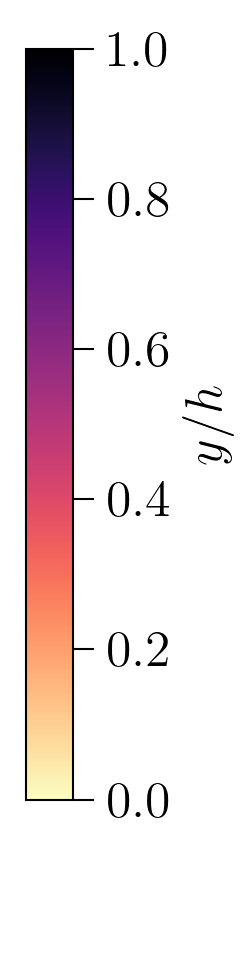}
         \end{subfigure}

        \caption{The streamwise energy spectra of the turbulent heat transfer $k_xE_{-T^\prime v^\prime}^x/\Delta TU_b h$ for the compliant-wall (left) and rigid-wall (right) cases at $Pr=7$. The line color indicates the different wall-normal distances from the bottom hot-wall (bright) to the channel center (dark).} 
        \label{fig: Ettx spectra fluc} 
\end{figure}

To quantify these observations, we look in figure \ref{fig: joint pdf} at the joint probability density function (or jpdf) of $u^\prime v^\prime$ and $T^\prime v^\prime$ at one of the wall-normal positions considered in figure \ref{fig: instantaneous 2d events}. The jpdf of $u^\prime v^\prime$ for the rigid-wall case has the classical distributions, with significant Q2 (ejection) and Q4 (sweep) events compared to Q1 and Q3, with a peak located in Q4. This distribution well relates to what qualitatively observed in figure \ref{fig: instantaneous 2d events}. A similar shape is found for the case with compliant walls, with however much wider distributions indicating stronger fluctuations. Also, the location of the peak is shifted towards the center of the diagram, which agrees with the report that increasing wall compliance leads to more important ejections \cite{Ardekani_Rosti_Brandt_2019}. The jpdf of $T^\prime v^\prime$ shows a rather uniform distribution for the case with rigid wall, slightly tilted in favour of Q1 and Q3 and with a peak in Q3. The distribution is mostly elongated in the vertical direction, indicating that the temperature fluctuations are relatively smaller than the velocity ones. This is consistent with what previously observed in figure \ref{fig: temp statistics}(right), which demonstrates limited temperature fluctuations at the selected wall-normal distance ($y\approx0.2h$). In the compliant-wall case, the jpdf of $T^\prime v^\prime$ becomes significantly wider than in the rigid-wall case, consistently with the enhanced heat transfer over the compliant walls. Note that, the distribution significantly expands along the horizontal axis compared to the rigid-wall case, indicating stronger temperature fluctuations in agreement with figure \ref{fig: heat flux}. For both the rigid and compliant cases, the predominance of Q2 and Q4 for $u^\prime v^\prime$ and of Q1 and Q3 for $T^\prime v^\prime$ confirms the importance of sweep and ejection events for transferring both momentum and heat in the wall-normal direction, with the most common transfers coming from the sweep events and the most intense ones from the ejections.

To investigate the scale-dependent variation of the convective flux, figure \ref{fig: Ettx spectra fluc} shows the streamwise one-dimensional premultiplied spectra of the turbulent heat transfer $k_xE^x_{-T^\prime v^\prime }$ for the rigid- and compliant-wall cases at $Pr=7$. The spectra of the compliant-wall case has a significantly larger magnitude of $-T^\prime v^\prime$ compared to the rigid-wall (almost double), consistently with the enhanced activity of the convective heat flux, as shown in figure \ref{fig: heat flux} (left). Furthermore, the energy peak for the compliant-wall is observed at a relatively smaller scale compared to that for the rigid-wall at low $y/h$, quantifying the previously observed fragmentation of the structures in the streamwise direction (shown in figure \ref{fig: instantaneous 2d events}) and providing a measure of the wavelength of the wall undulations along the streamwise direction. Note that, the peak wavelength and energy distributions of $k_xE^x_{-T^\prime v^\prime }$ for both cases show qualitatively similar trends to those of $k_xE^x_{-u^\prime v^\prime }$ (not shown here).

\begin{figure}[h]
     \centering
     \includegraphics[width=0.4\textwidth]{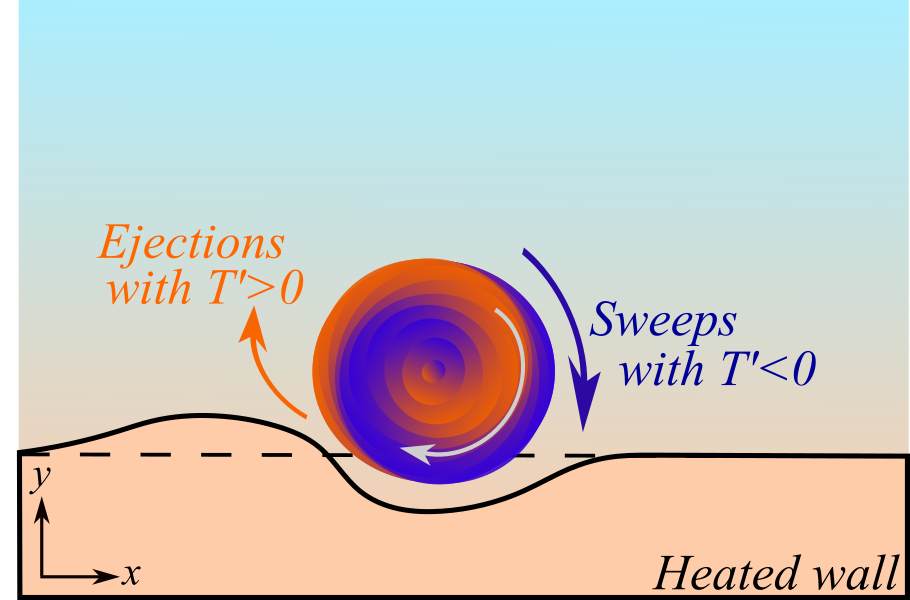}
     \caption{Conceptual sketch of the turbulent heat transfer alterations due to a compliant wall.}
    \label{fig: mix sketch}
\end{figure}

Overall, these observations highlight the significant changes caused by the compliant walls in the vertical component of the velocity fluctuations and in the temperature fields, which ultimately result in an enhanced turbulent heat transfer. Indeed, we can create the following visual picture (figure \ref{fig: mix sketch}): when the flow moves towards the compliant wall carrying cold fluid, it forces the wall deformation, pushing the hot wall around the sweep location up towards the bulk of the channel; when the wall restores back towards its equilibrium position, the fluid which ejected up into the bulk is now hot, having been heated by the surrounding hot walls.

\begin{figure}[tb]
     \centering
     \begin{subfigure}[c]{0.32\textwidth}
         \centering
         \includegraphics[width=1\textwidth]{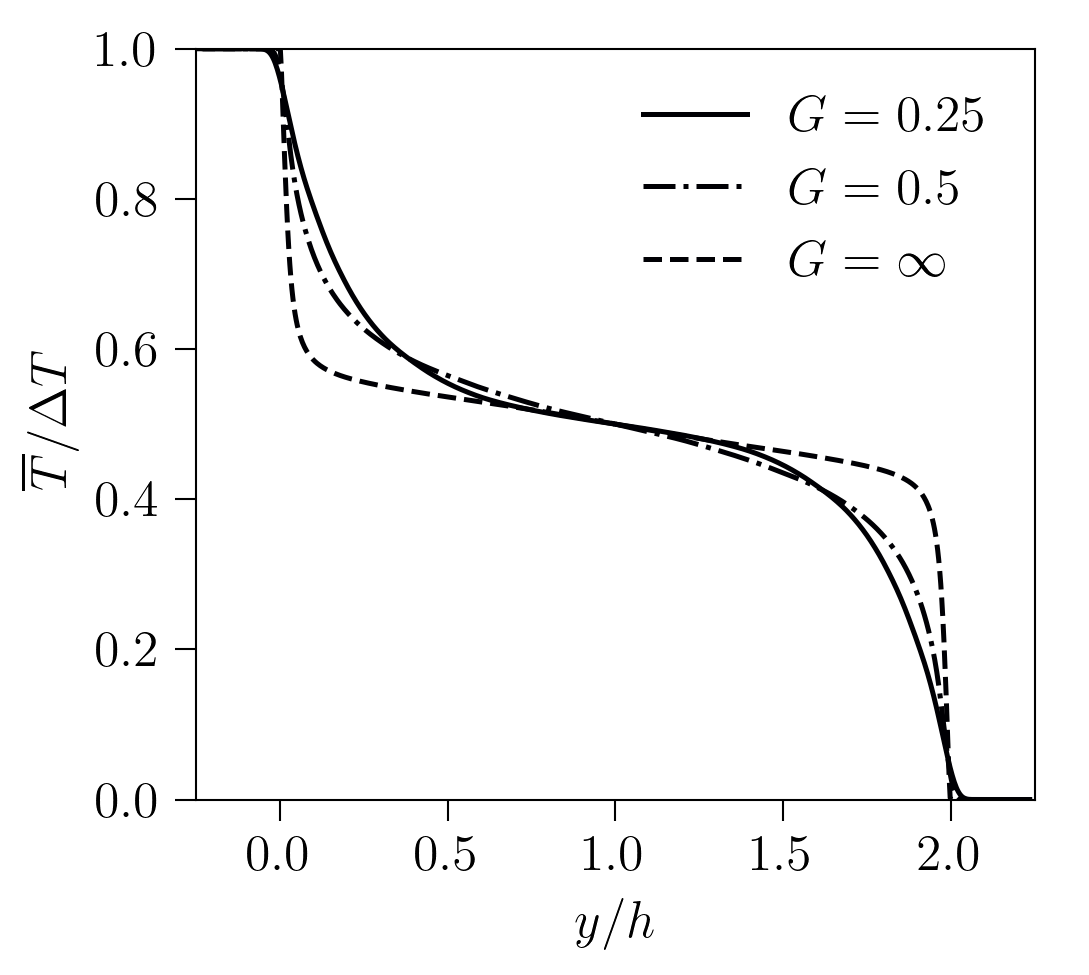}
         \end{subfigure}
     \hfill
     \begin{subfigure}[c]{0.32\textwidth}
         \centering
         \includegraphics[width=1\textwidth]{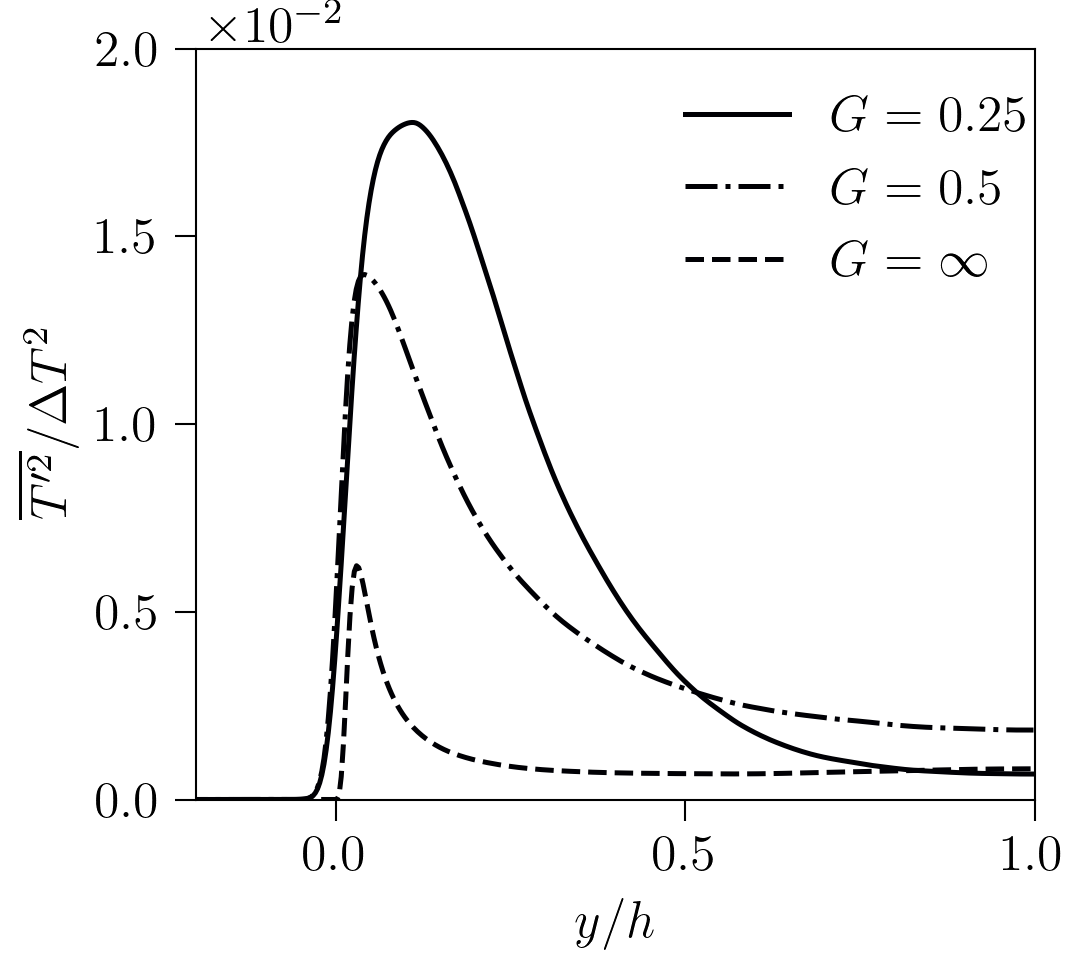}
         \end{subfigure}
     \hfill
     \begin{subfigure}[c]{0.32\textwidth}
         \centering
         \includegraphics[width=0.95\textwidth]{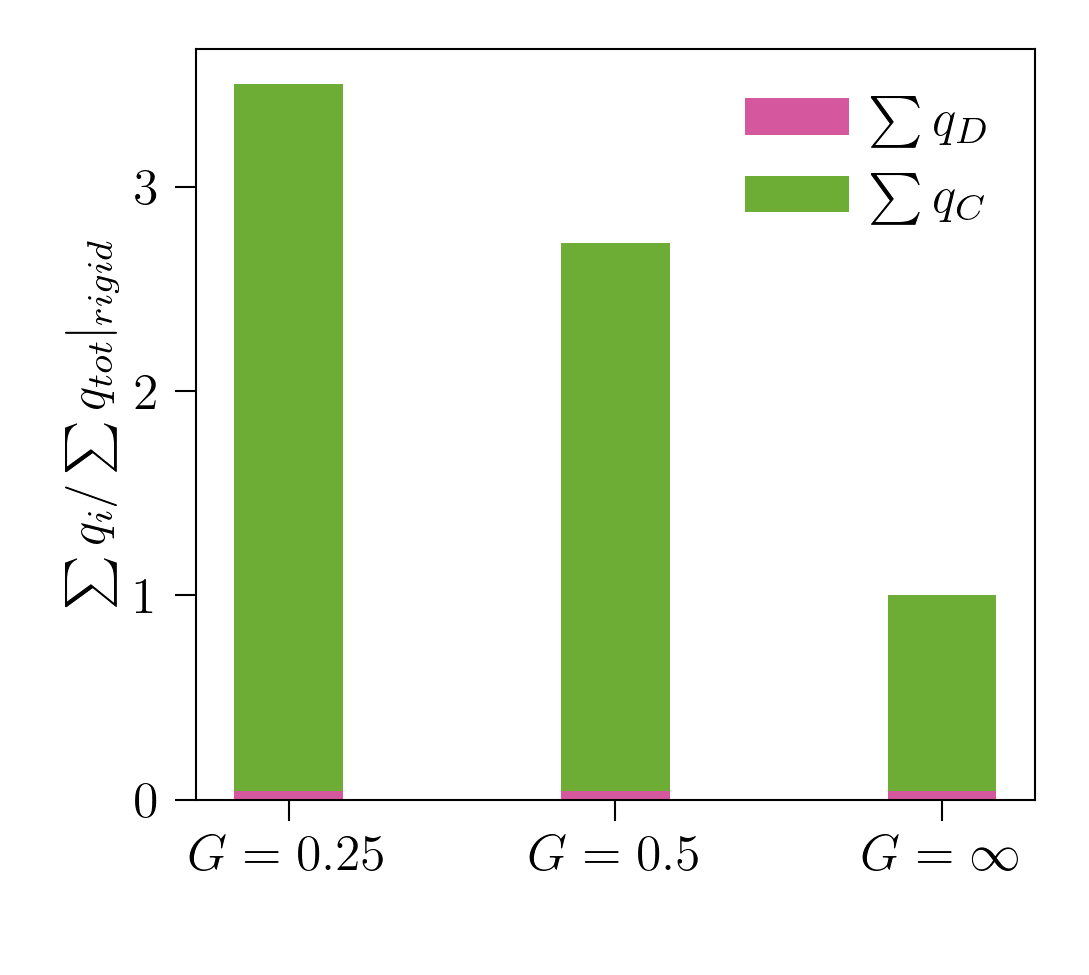}
         \end{subfigure}
        \caption{The wall-normal profiles of the thermal fields statistics for different transverse modulus at $Pr=7$: (a) mean temperature profiles and (b) variance of the temperature fluctuations. The line style represents the different transverse modulus of elasticity: $G=0.25$ (solid line), $G=0.5$ (dashed-dotted line), and $G=\infty$ (dashed line). (c) The integral contribution of the total diffusive $q_{D}$ (magenta) and convective $q_{C}$ (green) heat flux components introduced in Eq.~(\ref{eq: heat flux}) for different transverse modulus of elasticity.}
        \label{fig:varG}
\end{figure}

\subsection{The effect of the wall elasticity} \label{sec:b}
In this section, we study the effect of wall elasticity, i.e., different transversal modulus of elasticity $G$, by adding one additional intermediate value of elasticity $G=0.5 \rho U_b^2$ to the the two previously considered: $G=0.25\rho U_b^2$ (very deformable wall) and $G=\infty$ (rigid wall). Note that, while we have simulated all the Prandtl numbers considered in this study, in the following we show only those at $Pr=7$, since no qualitative difference can be found for the other values. As previously shown, when the wall is elastic, not only the turbulent flow statistics but also the heat transfer statistics tend to intensify due to the increased wall motion. Here, we want to assess if this is a continuous increase or if different regimes arise.  Note that, the two values of the elastic shear modulus have been selected to be in the two different regimes of wall oscillations identified in Ref.~\cite{Rosti_Brandt_2017,wang2020interaction}, one dominated by streamwise aligned ridges and the other with highly coherent spanwise oscillations.

Figures \ref{fig:varG}(a) and (b) demonstrate that the near-wall temperature gradient becomes progressively smaller when $G$ reduces, accompanied by an increase of the peak of the fluctuations which shifts away from the wall. As done before, we quantify the heat flux balance in figure \ref{fig:varG}(c) by showing the diffusive and convective contributions integrated across the channel height. We confirm once more that the convective term is dominant over the viscous one, and significantly increases when the wall is let free to move and adapt to the flow. All the three panels of the figure provide a coherent picture, in which the temperature and heat flux statistics are quite close for the two different elasticity considered, strongly departing from that of the rigid wall. This is found notwithstanding the fact that the two wall oscillations are significantly differ, with a root mean square values of wall oscillation equal to $0.025h$ for $G=0.25 \rho U_b^2$ and reducing to $0.015h$ for $G=0.5 \rho U_b^2$. Thus, while the wall oscillations and consequent momentum transfer are in two different regimes, the resulting heat transfer modifications are roughly the same. Thus, the results suggest that even a small level of wall compliance is able to bring the modifications in the heat transfer.

\subsection{Heat and momentum transfer dissimilarity} \label{sec:c}

\begin{figure}[h!]
     \centering
     \includegraphics[width=0.5\textwidth]{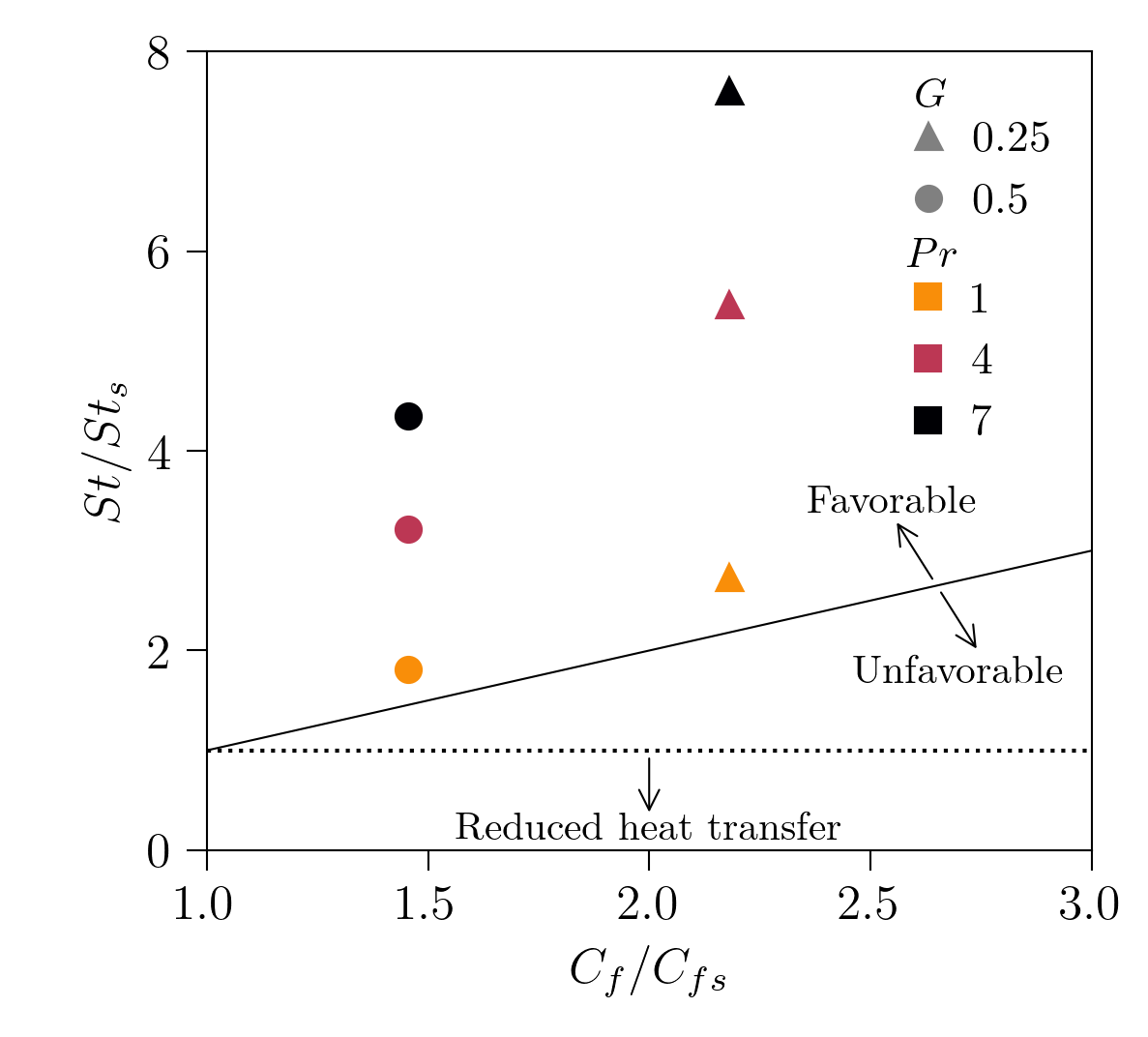}
     \caption{Reynolds analogy plot for all considered compliant-wall cases. The symbols represent different wall elasticity: $\triangle$ for $G/\rho U_b^2 =0.25$, and $\bigcirc$ for $G/\rho U_b^2 =0.5$. The line colors indicate the different Prandtl number: $Pr=1$ (orange), $Pr=4$ (magenta), and $Pr=7$ (black). The thin line marks $St/St_s = C_f/C_{fs}$.}
    \label{fig: ReynoldsAnalogy}
\end{figure}

In this last section, we investigate the breakdown of the Reynolds analogy in the compliant-wall cases, i.e., the similarity between the heat and momentum transfer. The breakdown is estimated through the ratio of the fractional increase of the Stanton number $St$ and skin friction coefficient $C_f$ in the compliant-wall case, with respect to the rigid-wall, where the Stanton number represents the non-dimensional wall heat flux, and the skin-friction coefficient represents the non-dimensional wall shear stress $\tau_w$, here defined as
\begin{equation}
    St =\frac{q_{tot}}{\rho U_b c_p}, \quad \textrm{and} \quad C_f =\frac{2\tau_w}{\rho U_b^2},
    \label{eq: def St and Cf}
\end{equation}
where $c_p$ is the specific heat capacity, assumed as unity. The deviation from the Reynolds analogy can be categorized into two types, related to the heat transfer efficiency: favorable, when the increase of $St$ is greater than $C_f$, and unfavorable, when the increase of $St$ is less than $C_f$ \cite{Rouhi_Endrikat_Modesti_Sandberg_Oda_Tanimoto_Hutchins_Chung_2022}.
To this end, figure \ref{fig: ReynoldsAnalogy} plots the Stanton number versus the skin friction coefficient of the compliant-wall cases, normalized by those for the rigid-smooth wall, and shows the breakdown of the Reynolds analogy in the present cases. All symbols are distributed in the $St/St_s > C_f/C_{fs}$ region, representing a favorable breakdown and efficient heat transfer. Interestingly, the favorable breakdown of the heat transfer over compliant walls is distinct from the previously observed unfavorable breakdowns in the turbulent heat transfer over both rough \cite{Rouhi_Endrikat_Modesti_Sandberg_Oda_Tanimoto_Hutchins_Chung_2022} and porous walls \cite{khorasani2025porous}.

\section{Conclusion} \label{sec:con}
In this study we investigate the effects of a compliant wall on turbulent flow mixing, specifically its passive scalar heat transfer. To do so, we perform direct numerical simulations of turbulent channel flows bounded by two compliant walls, and investigate the effect on the temperature field described by a passive advection-diffusion equation. We perform a parametric study, where we vary the wall elasticity and scalar diffusivity, to examine their effects on the turbulent heat transfer. Overall, the results show that the heat transfer is strongly increased by the wall elasticity, driven by an enhanced turbulent convection. 

We show that, when wall flexibility increases leading to enhanced wall motion, not only the momentum transfer \cite{Rosti_Brandt_2017, Ardekani_Rosti_Brandt_2019, Esteghamatian2022spatiotemporal, Koseki_Aswathy_Rosti_2025}, but also the heat transfer is strengthened. Consistent with the previous studies, we find that the momentum transfer is intensified with the compliant wall by the wall oscillations which cause strong velocity fluctuations, especially those in the wall-normal direction. In the thermal field, the effects of the compliant-wall are clearly observed in the mean and variance of the temperature fluctuations. Indeed, the cases with compliant walls exhibit a reduced temperature gradient near the wall and an increased temperature variance across the whole channel. When examining the heat transfer balance, our results show that, while the diffusive heat transfer is reduced in the presence of a compliant wall, this is more than compensated by a strong enhancement of the convective term, which dominates even at the wall. Note that, in the case of a compliant wall, this term comes from both the fluid and solid phase, since the wall itself oscillates in the wall-normal direction. We explain the enhanced convective flux by looking the temperature-velocity correlation. The simulations show that the enhanced temperature fluctuations correlate well with the enhanced wall-normal fluctuations, which are due to intensified sweep and ejection event in the flow. We can interpret this as a consequence of the non-zero wall-normal velocity fluctuations that bring cold fluid towards the wall and release the hot fluid into the bulk region when the wall oscillates.

We also investigate the effects of varying Prandtl number and wall elasticity on the turbulent heat transfer over compliant walls. On the one hand, the variation of the Prandtl number carries similar modifications for both the rigid and compliant cases. On the other hand, wall elasticity has significant effects, with even a small level of wall compliance which is able to completely modify the heat transfer process, when compared to the case over rigid wall.

The present study has shown that heat transfer, and turbulent mixing more in general, can be significantly altered in the presence of an compliant wall. A mechanistic explanation for the modifications is that when the turbulent fluctuations sweep towards the wall carrying cold fluid from the bulk of the channel, they force the wall to indent and to raise the wall around upwards due to incompressibility. This has a double effect: first, the hot wall up in the bulk of the channel warms the colder fluid around, and second, when the wall restores back towards its equilibrium position, it ejects the hot fluid occupying its position. This heat transfer mechanism due the wall compliancy produces a favorable increase of the heat transfer compared to the  momentum one, thus promoting its possible exploitation in applications; indeed these results have various possible implications and applications, ranging from food processing to chemical reactors, where rapid mixing is highly desirable.

\acknowledgments
The research was supported by the Okinawa Institute of Science and Technology Graduate University (OIST) with subsidy funding to M.E.R. from the Cabinet Office, Government of Japan. M.E.R.~also acknowledges funding from the Japan Society for the Promotion of Science (JSPS), grants 24K00810 and 24K17210. The authors acknowledge the computer time provided by the Scientific Computing and Data Analysis section of the Core Facilities at OIST and the computational resources offered by the HPCI System Research Project with grants hp240006 and hp250035. 


\begin{thebibliography}{67}%
\makeatletter
\providecommand \@ifxundefined [1]{%
 \@ifx{#1\undefined}
}%
\providecommand \@ifnum [1]{%
 \ifnum #1\expandafter \@firstoftwo
 \else \expandafter \@secondoftwo
 \fi
}%
\providecommand \@ifx [1]{%
 \ifx #1\expandafter \@firstoftwo
 \else \expandafter \@secondoftwo
 \fi
}%
\providecommand \natexlab [1]{#1}%
\providecommand \enquote  [1]{``#1''}%
\providecommand \bibnamefont  [1]{#1}%
\providecommand \bibfnamefont [1]{#1}%
\providecommand \citenamefont [1]{#1}%
\providecommand \href@noop [0]{\@secondoftwo}%
\providecommand \href [0]{\begingroup \@sanitize@url \@href}%
\providecommand \@href[1]{\@@startlink{#1}\@@href}%
\providecommand \@@href[1]{\endgroup#1\@@endlink}%
\providecommand \@sanitize@url [0]{\catcode `\\12\catcode `\$12\catcode
  `\&12\catcode `\#12\catcode `\^12\catcode `\_12\catcode `\%12\relax}%
\providecommand \@@startlink[1]{}%
\providecommand \@@endlink[0]{}%
\providecommand \url  [0]{\begingroup\@sanitize@url \@url }%
\providecommand \@url [1]{\endgroup\@href {#1}{\urlprefix }}%
\providecommand \urlprefix  [0]{URL }%
\providecommand \Eprint [0]{\href }%
\providecommand \doibase [0]{https://doi.org/}%
\providecommand \selectlanguage [0]{\@gobble}%
\providecommand \bibinfo  [0]{\@secondoftwo}%
\providecommand \bibfield  [0]{\@secondoftwo}%
\providecommand \translation [1]{[#1]}%
\providecommand \BibitemOpen [0]{}%
\providecommand \bibitemStop [0]{}%
\providecommand \bibitemNoStop [0]{.\EOS\space}%
\providecommand \EOS [0]{\spacefactor3000\relax}%
\providecommand \BibitemShut  [1]{\csname bibitem#1\endcsname}%
\let\auto@bib@innerbib\@empty
\bibitem [{\citenamefont {Riley}\ \emph {et~al.}(1988)\citenamefont {Riley},
  \citenamefont {Gad-el Hak},\ and\ \citenamefont
  {Metcalfe}}]{riley1988complaint}%
  \BibitemOpen
  \bibfield  {author} {\bibinfo {author} {\bibfnamefont {J.~J.}\ \bibnamefont
  {Riley}}, \bibinfo {author} {\bibfnamefont {M.}~\bibnamefont {Gad-el Hak}},\
  and\ \bibinfo {author} {\bibfnamefont {R.~W.}\ \bibnamefont {Metcalfe}},\
  }\bibfield  {title} {\bibinfo {title} {Complaint coatings},\ }\href@noop {}
  {\bibfield  {journal} {\bibinfo  {journal} {Annual Review of Fluid
  Mechanics}\ }\textbf {\bibinfo {volume} {20}},\ \bibinfo {pages} {393}
  (\bibinfo {year} {1988})}\BibitemShut {NoStop}%
\bibitem [{\citenamefont {Carpenter}(1993)}]{carpenter1993optimization}%
  \BibitemOpen
  \bibfield  {author} {\bibinfo {author} {\bibfnamefont {P.~W.}\ \bibnamefont
  {Carpenter}},\ }\bibfield  {title} {\bibinfo {title} {Optimization of
  multiple-panel compliant walls for delay of laminar-turbulent transition},\
  }\href@noop {} {\bibfield  {journal} {\bibinfo  {journal} {Aiaa Journal}\
  }\textbf {\bibinfo {volume} {31}},\ \bibinfo {pages} {1187} (\bibinfo {year}
  {1993})}\BibitemShut {NoStop}%
\bibitem [{\citenamefont {Nagy}\ \emph {et~al.}(2022)\citenamefont {Nagy},
  \citenamefont {Szab{\'o}},\ and\ \citenamefont {Paal}}]{nagy2022effect}%
  \BibitemOpen
  \bibfield  {author} {\bibinfo {author} {\bibfnamefont {P.~T.}\ \bibnamefont
  {Nagy}}, \bibinfo {author} {\bibfnamefont {A.}~\bibnamefont {Szab{\'o}}},\
  and\ \bibinfo {author} {\bibfnamefont {G.}~\bibnamefont {Paal}},\ }\bibfield
  {title} {\bibinfo {title} {The effect of spanwise and streamwise elastic
  coating on boundary layer transition},\ }\href@noop {} {\bibfield  {journal}
  {\bibinfo  {journal} {Journal of Fluids and Structures}\ }\textbf {\bibinfo
  {volume} {110}},\ \bibinfo {pages} {103521} (\bibinfo {year}
  {2022})}\BibitemShut {NoStop}%
\bibitem [{\citenamefont {Semenov}(1991)}]{semenov1991conditions}%
  \BibitemOpen
  \bibfield  {author} {\bibinfo {author} {\bibfnamefont {B.}~\bibnamefont
  {Semenov}},\ }\bibfield  {title} {\bibinfo {title} {On conditions of
  modelling and choice of viscoelastic coatings for drag reduction},\ }in\
  \href@noop {} {\emph {\bibinfo {booktitle} {Recent developments in turbulence
  management}}}\ (\bibinfo  {publisher} {Springer},\ \bibinfo {year} {1991})\
  pp.\ \bibinfo {pages} {241--262}\BibitemShut {NoStop}%
\bibitem [{\citenamefont {Carpenter}\ \emph {et~al.}(2000)\citenamefont
  {Carpenter}, \citenamefont {Davies},\ and\ \citenamefont
  {Lucey}}]{carpenter2000hydrodynamics}%
  \BibitemOpen
  \bibfield  {author} {\bibinfo {author} {\bibfnamefont {P.~W.}\ \bibnamefont
  {Carpenter}}, \bibinfo {author} {\bibfnamefont {C.}~\bibnamefont {Davies}},\
  and\ \bibinfo {author} {\bibfnamefont {A.~D.}\ \bibnamefont {Lucey}},\
  }\bibfield  {title} {\bibinfo {title} {Hydrodynamics and compliant walls:
  Does the dolphin have a secret?},\ }\href@noop {} {\bibfield  {journal}
  {\bibinfo  {journal} {Current Science}\ ,\ \bibinfo {pages} {758}} (\bibinfo
  {year} {2000})}\BibitemShut {NoStop}%
\bibitem [{\citenamefont {Gad-el Hak}(2002)}]{gad2002compliant}%
  \BibitemOpen
  \bibfield  {author} {\bibinfo {author} {\bibfnamefont {M.}~\bibnamefont
  {Gad-el Hak}},\ }\bibfield  {title} {\bibinfo {title} {Compliant coatings for
  drag reduction},\ }\href@noop {} {\bibfield  {journal} {\bibinfo  {journal}
  {Progress in Aerospace Sciences}\ }\textbf {\bibinfo {volume} {38}},\
  \bibinfo {pages} {77} (\bibinfo {year} {2002})}\BibitemShut {NoStop}%
\bibitem [{\citenamefont {Nisewanger}(1964)}]{nisewanger1964flow}%
  \BibitemOpen
  \bibfield  {author} {\bibinfo {author} {\bibfnamefont {C.}~\bibnamefont
  {Nisewanger}},\ }\bibfield  {title} {\bibinfo {title} {Flow noise and drag
  measurements of vehicle with compliant coating},\ }\href@noop {} {\bibfield
  {journal} {\bibinfo  {journal} {NAVWEPS Rep}\ }\textbf {\bibinfo {volume}
  {8518}} (\bibinfo {year} {1964})}\BibitemShut {NoStop}%
\bibitem [{\citenamefont {Kumaran}(2021)}]{kumaran2021stability}%
  \BibitemOpen
  \bibfield  {author} {\bibinfo {author} {\bibfnamefont {V.}~\bibnamefont
  {Kumaran}},\ }\bibfield  {title} {\bibinfo {title} {Stability and the
  transition to turbulence in the flow through conduits with compliant walls},\
  }\href@noop {} {\bibfield  {journal} {\bibinfo  {journal} {Journal of Fluid
  Mechanics}\ }\textbf {\bibinfo {volume} {924}},\ \bibinfo {pages} {P1}
  (\bibinfo {year} {2021})}\BibitemShut {NoStop}%
\bibitem [{\citenamefont {Benjamin}(1960)}]{benjamin1960effects}%
  \BibitemOpen
  \bibfield  {author} {\bibinfo {author} {\bibfnamefont {T.~B.}\ \bibnamefont
  {Benjamin}},\ }\bibfield  {title} {\bibinfo {title} {Effects of a flexible
  boundary on hydrodynamic stability},\ }\href@noop {} {\bibfield  {journal}
  {\bibinfo  {journal} {Journal of Fluid Mechanics}\ }\textbf {\bibinfo
  {volume} {9}},\ \bibinfo {pages} {513} (\bibinfo {year} {1960})}\BibitemShut
  {NoStop}%
\bibitem [{\citenamefont {Benjamin}(1963)}]{benjamin1963threefold}%
  \BibitemOpen
  \bibfield  {author} {\bibinfo {author} {\bibfnamefont {T.~B.}\ \bibnamefont
  {Benjamin}},\ }\bibfield  {title} {\bibinfo {title} {The threefold
  classification of unstable disturbances in flexible surfaces bounding
  inviscid flows},\ }\href@noop {} {\bibfield  {journal} {\bibinfo  {journal}
  {Journal of Fluid Mechanics}\ }\textbf {\bibinfo {volume} {16}},\ \bibinfo
  {pages} {436} (\bibinfo {year} {1963})}\BibitemShut {NoStop}%
\bibitem [{\citenamefont {Landahl}(1962)}]{landahl1962stability}%
  \BibitemOpen
  \bibfield  {author} {\bibinfo {author} {\bibfnamefont {M.~T.}\ \bibnamefont
  {Landahl}},\ }\bibfield  {title} {\bibinfo {title} {On the stability of a
  laminar incompressible boundary layer over a flexible surface},\ }\href@noop
  {} {\bibfield  {journal} {\bibinfo  {journal} {Journal of Fluid Mechanics}\
  }\textbf {\bibinfo {volume} {13}},\ \bibinfo {pages} {609} (\bibinfo {year}
  {1962})}\BibitemShut {NoStop}%
\bibitem [{\citenamefont {Carpenter}\ and\ \citenamefont
  {Garrad}(1985)}]{carpenter1985hydrodynamic}%
  \BibitemOpen
  \bibfield  {author} {\bibinfo {author} {\bibfnamefont {P.}~\bibnamefont
  {Carpenter}}\ and\ \bibinfo {author} {\bibfnamefont {A.}~\bibnamefont
  {Garrad}},\ }\bibfield  {title} {\bibinfo {title} {The hydrodynamic stability
  of flow over kramer-type compliant surfaces. part 1. tollmien-schlichting
  instabilities},\ }\href@noop {} {\bibfield  {journal} {\bibinfo  {journal}
  {Journal of Fluid Mechanics}\ }\textbf {\bibinfo {volume} {155}},\ \bibinfo
  {pages} {465} (\bibinfo {year} {1985})}\BibitemShut {NoStop}%
\bibitem [{\citenamefont {Carpenter}\ and\ \citenamefont
  {Garrad}(1986)}]{carpenter1986hydrodynamic}%
  \BibitemOpen
  \bibfield  {author} {\bibinfo {author} {\bibfnamefont {P.}~\bibnamefont
  {Carpenter}}\ and\ \bibinfo {author} {\bibfnamefont {A.}~\bibnamefont
  {Garrad}},\ }\bibfield  {title} {\bibinfo {title} {The hydrodynamic stability
  of flow over kramer-type compliant surfaces. part 2. flow-induced surface
  instabilities},\ }\href@noop {} {\bibfield  {journal} {\bibinfo  {journal}
  {Journal of Fluid Mechanics}\ }\textbf {\bibinfo {volume} {170}},\ \bibinfo
  {pages} {199} (\bibinfo {year} {1986})}\BibitemShut {NoStop}%
\bibitem [{\citenamefont {Rotenberry}\ and\ \citenamefont
  {Saffman}(1990)}]{rotenberry1990effect}%
  \BibitemOpen
  \bibfield  {author} {\bibinfo {author} {\bibfnamefont {J.~M.}\ \bibnamefont
  {Rotenberry}}\ and\ \bibinfo {author} {\bibfnamefont {P.~G.}\ \bibnamefont
  {Saffman}},\ }\bibfield  {title} {\bibinfo {title} {Effect of compliant
  boundaries on weakly nonlinear shear waves in channel flow},\ }\href@noop {}
  {\bibfield  {journal} {\bibinfo  {journal} {SIAM Journal on Applied
  Mathematics}\ }\textbf {\bibinfo {volume} {50}},\ \bibinfo {pages} {361}
  (\bibinfo {year} {1990})}\BibitemShut {NoStop}%
\bibitem [{\citenamefont {Rotenberry}(1992)}]{rotenberry1992finite}%
  \BibitemOpen
  \bibfield  {author} {\bibinfo {author} {\bibfnamefont {J.~M.}\ \bibnamefont
  {Rotenberry}},\ }\bibfield  {title} {\bibinfo {title} {Finite-amplitude shear
  waves in a channel with compliant boundaries},\ }\href@noop {} {\bibfield
  {journal} {\bibinfo  {journal} {Physics of Fluids A: Fluid Dynamics}\
  }\textbf {\bibinfo {volume} {4}},\ \bibinfo {pages} {270} (\bibinfo {year}
  {1992})}\BibitemShut {NoStop}%
\bibitem [{\citenamefont {Davies}\ and\ \citenamefont
  {Carpenter}(997b)}]{davies1997numerical}%
  \BibitemOpen
  \bibfield  {author} {\bibinfo {author} {\bibfnamefont {C.}~\bibnamefont
  {Davies}}\ and\ \bibinfo {author} {\bibfnamefont {P.~W.}\ \bibnamefont
  {Carpenter}},\ }\bibfield  {title} {\bibinfo {title} {Numerical simulation of
  the evolution of tollmien--schlichting waves over finite compliant panels},\
  }\href@noop {} {\bibfield  {journal} {\bibinfo  {journal} {Journal of Fluid
  Mechanics}\ }\textbf {\bibinfo {volume} {335}},\ \bibinfo {pages} {361}
  (\bibinfo {year} {1997b})}\BibitemShut {NoStop}%
\bibitem [{\citenamefont {Davies}\ and\ \citenamefont
  {Carpenter}(997a)}]{davies1997instabilities}%
  \BibitemOpen
  \bibfield  {author} {\bibinfo {author} {\bibfnamefont {C.}~\bibnamefont
  {Davies}}\ and\ \bibinfo {author} {\bibfnamefont {P.~W.}\ \bibnamefont
  {Carpenter}},\ }\bibfield  {title} {\bibinfo {title} {Instabilities in a
  plane channel flow between compliant walls},\ }\href@noop {} {\bibfield
  {journal} {\bibinfo  {journal} {Journal of Fluid Mechanics}\ }\textbf
  {\bibinfo {volume} {352}},\ \bibinfo {pages} {205} (\bibinfo {year}
  {1997a})}\BibitemShut {NoStop}%
\bibitem [{\citenamefont {Duncan}(1986)}]{duncan1986response}%
  \BibitemOpen
  \bibfield  {author} {\bibinfo {author} {\bibfnamefont {J.~H.}\ \bibnamefont
  {Duncan}},\ }\bibfield  {title} {\bibinfo {title} {The response of an
  incompressible, viscoelastic coating to pressure fluctuations in a turbulent
  boundary layer},\ }\href@noop {} {\bibfield  {journal} {\bibinfo  {journal}
  {Journal of Fluid Mechanics}\ }\textbf {\bibinfo {volume} {171}},\ \bibinfo
  {pages} {339} (\bibinfo {year} {1986})}\BibitemShut {NoStop}%
\bibitem [{\citenamefont {Benschop}\ \emph {et~al.}(2019)\citenamefont
  {Benschop}, \citenamefont {Greidanus}, \citenamefont {Delfos}, \citenamefont
  {Westerweel},\ and\ \citenamefont {Breugem}}]{benschop2019deformation}%
  \BibitemOpen
  \bibfield  {author} {\bibinfo {author} {\bibfnamefont {H.}~\bibnamefont
  {Benschop}}, \bibinfo {author} {\bibfnamefont {A.}~\bibnamefont {Greidanus}},
  \bibinfo {author} {\bibfnamefont {R.}~\bibnamefont {Delfos}}, \bibinfo
  {author} {\bibfnamefont {J.}~\bibnamefont {Westerweel}},\ and\ \bibinfo
  {author} {\bibfnamefont {W.-P.}\ \bibnamefont {Breugem}},\ }\bibfield
  {title} {\bibinfo {title} {Deformation of a linear viscoelastic compliant
  coating in a turbulent flow},\ }\href@noop {} {\bibfield  {journal} {\bibinfo
   {journal} {Journal of Fluid Mechanics}\ }\textbf {\bibinfo {volume} {859}},\
  \bibinfo {pages} {613} (\bibinfo {year} {2019})}\BibitemShut {NoStop}%
\bibitem [{\citenamefont {Zhang}\ \emph {et~al.}(2015)\citenamefont {Zhang},
  \citenamefont {Miorini},\ and\ \citenamefont {Katz}}]{zhang2015integrating}%
  \BibitemOpen
  \bibfield  {author} {\bibinfo {author} {\bibfnamefont {C.}~\bibnamefont
  {Zhang}}, \bibinfo {author} {\bibfnamefont {R.}~\bibnamefont {Miorini}},\
  and\ \bibinfo {author} {\bibfnamefont {J.}~\bibnamefont {Katz}},\ }\bibfield
  {title} {\bibinfo {title} {Integrating mach--zehnder interferometry with tpiv
  to measure the time-resolved deformation of a compliant wall along with the
  3d velocity field in a turbulent channel flow},\ }\href@noop {} {\bibfield
  {journal} {\bibinfo  {journal} {Experiments in Fluids}\ }\textbf {\bibinfo
  {volume} {56}},\ \bibinfo {pages} {1} (\bibinfo {year} {2015})}\BibitemShut
  {NoStop}%
\bibitem [{\citenamefont {Zhang}\ \emph {et~al.}(2017)\citenamefont {Zhang},
  \citenamefont {Wang}, \citenamefont {Blake},\ and\ \citenamefont
  {Katz}}]{zhang2017deformation}%
  \BibitemOpen
  \bibfield  {author} {\bibinfo {author} {\bibfnamefont {C.}~\bibnamefont
  {Zhang}}, \bibinfo {author} {\bibfnamefont {J.}~\bibnamefont {Wang}},
  \bibinfo {author} {\bibfnamefont {W.}~\bibnamefont {Blake}},\ and\ \bibinfo
  {author} {\bibfnamefont {J.}~\bibnamefont {Katz}},\ }\bibfield  {title}
  {\bibinfo {title} {Deformation of a compliant wall in a turbulent channel
  flow},\ }\href@noop {} {\bibfield  {journal} {\bibinfo  {journal} {Journal of
  Fluid Mechanics}\ }\textbf {\bibinfo {volume} {823}},\ \bibinfo {pages} {345}
  (\bibinfo {year} {2017})}\BibitemShut {NoStop}%
\bibitem [{\citenamefont {Wang}\ \emph {et~al.}(2020)\citenamefont {Wang},
  \citenamefont {Koley},\ and\ \citenamefont {Katz}}]{wang2020interaction}%
  \BibitemOpen
  \bibfield  {author} {\bibinfo {author} {\bibfnamefont {J.}~\bibnamefont
  {Wang}}, \bibinfo {author} {\bibfnamefont {S.~S.}\ \bibnamefont {Koley}},\
  and\ \bibinfo {author} {\bibfnamefont {J.}~\bibnamefont {Katz}},\ }\bibfield
  {title} {\bibinfo {title} {On the interaction of a compliant wall with a
  turbulent boundary layer},\ }\href@noop {} {\bibfield  {journal} {\bibinfo
  {journal} {Journal of Fluid Mechanics}\ }\textbf {\bibinfo {volume} {899}},\
  \bibinfo {pages} {A20} (\bibinfo {year} {2020})}\BibitemShut {NoStop}%
\bibitem [{\citenamefont {Greidanus}\ \emph {et~al.}(2022)\citenamefont
  {Greidanus}, \citenamefont {Delfos}, \citenamefont {Picken},\ and\
  \citenamefont {Westerweel}}]{greidanus2022response}%
  \BibitemOpen
  \bibfield  {author} {\bibinfo {author} {\bibfnamefont {A.}~\bibnamefont
  {Greidanus}}, \bibinfo {author} {\bibfnamefont {R.}~\bibnamefont {Delfos}},
  \bibinfo {author} {\bibfnamefont {S.}~\bibnamefont {Picken}},\ and\ \bibinfo
  {author} {\bibfnamefont {J.}~\bibnamefont {Westerweel}},\ }\bibfield  {title}
  {\bibinfo {title} {Response regimes in the fluid-structure interaction of
  wall turbulence over a compliant coating},\ }\href@noop {} {\bibfield
  {journal} {\bibinfo  {journal} {Journal of Fluid Mechanics}\ }\textbf
  {\bibinfo {volume} {952}},\ \bibinfo {pages} {A1} (\bibinfo {year}
  {2022})}\BibitemShut {NoStop}%
\bibitem [{\citenamefont {Endo}\ and\ \citenamefont
  {Himeno}(2002)}]{endo2002direct}%
  \BibitemOpen
  \bibfield  {author} {\bibinfo {author} {\bibfnamefont {T.}~\bibnamefont
  {Endo}}\ and\ \bibinfo {author} {\bibfnamefont {R.}~\bibnamefont {Himeno}},\
  }\bibfield  {title} {\bibinfo {title} {Direct numerical simulation of
  turbulent flow over a compliant surface},\ }\href@noop {} {\bibfield
  {journal} {\bibinfo  {journal} {Journal of Turbulence}\ }\textbf {\bibinfo
  {volume} {3}},\ \bibinfo {pages} {007} (\bibinfo {year} {2002})}\BibitemShut
  {NoStop}%
\bibitem [{\citenamefont {Kim}\ and\ \citenamefont
  {Choi}(2014)}]{kim2014space}%
  \BibitemOpen
  \bibfield  {author} {\bibinfo {author} {\bibfnamefont {E.}~\bibnamefont
  {Kim}}\ and\ \bibinfo {author} {\bibfnamefont {H.}~\bibnamefont {Choi}},\
  }\bibfield  {title} {\bibinfo {title} {Space--time characteristics of a
  compliant wall in a turbulent channel flow},\ }\href@noop {} {\bibfield
  {journal} {\bibinfo  {journal} {Journal of fluid mechanics}\ }\textbf
  {\bibinfo {volume} {756}},\ \bibinfo {pages} {30} (\bibinfo {year}
  {2014})}\BibitemShut {NoStop}%
\bibitem [{\citenamefont {Rosti}\ and\ \citenamefont
  {Brandt}(2017)}]{Rosti_Brandt_2017}%
  \BibitemOpen
  \bibfield  {author} {\bibinfo {author} {\bibfnamefont {M.~E.}\ \bibnamefont
  {Rosti}}\ and\ \bibinfo {author} {\bibfnamefont {L.}~\bibnamefont {Brandt}},\
  }\bibfield  {title} {\bibinfo {title} {Numerical simulation of turbulent
  channel flow over a viscous hyper-elastic wall},\ }\href
  {https://doi.org/10.1017/jfm.2017.617} {\bibfield  {journal} {\bibinfo
  {journal} {Journal of Fluid Mechanics}\ }\textbf {\bibinfo {volume} {830}},\
  \bibinfo {pages} {708–735} (\bibinfo {year} {2017})}\BibitemShut {NoStop}%
\bibitem [{\citenamefont {Ardekani}\ \emph {et~al.}(2019)\citenamefont
  {Ardekani}, \citenamefont {Rosti},\ and\ \citenamefont
  {Brandt}}]{Ardekani_Rosti_Brandt_2019}%
  \BibitemOpen
  \bibfield  {author} {\bibinfo {author} {\bibfnamefont {M.~N.}\ \bibnamefont
  {Ardekani}}, \bibinfo {author} {\bibfnamefont {M.~E.}\ \bibnamefont
  {Rosti}},\ and\ \bibinfo {author} {\bibfnamefont {L.}~\bibnamefont
  {Brandt}},\ }\bibfield  {title} {\bibinfo {title} {Turbulent flow of
  finite-size spherical particles in channels with viscous hyper-elastic
  walls},\ }\href {https://doi.org/10.1017/jfm.2019.413} {\bibfield  {journal}
  {\bibinfo  {journal} {Journal of Fluid Mechanics}\ }\textbf {\bibinfo
  {volume} {873}},\ \bibinfo {pages} {410–440} (\bibinfo {year}
  {2019})}\BibitemShut {NoStop}%
\bibitem [{\citenamefont {Esteghamatian}\ \emph {et~al.}(2022)\citenamefont
  {Esteghamatian}, \citenamefont {Katz},\ and\ \citenamefont
  {Zaki}}]{Esteghamatian2022spatiotemporal}%
  \BibitemOpen
  \bibfield  {author} {\bibinfo {author} {\bibfnamefont {A.}~\bibnamefont
  {Esteghamatian}}, \bibinfo {author} {\bibfnamefont {J.}~\bibnamefont
  {Katz}},\ and\ \bibinfo {author} {\bibfnamefont {T.~A.}\ \bibnamefont
  {Zaki}},\ }\bibfield  {title} {\bibinfo {title} {Spatiotemporal
  characterization of turbulent channel flow with a hyperelastic compliant
  wall},\ }\href@noop {} {\bibfield  {journal} {\bibinfo  {journal} {Journal of
  Fluid Mechanics}\ }\textbf {\bibinfo {volume} {942}},\ \bibinfo {pages} {A35}
  (\bibinfo {year} {2022})}\BibitemShut {NoStop}%
\bibitem [{\citenamefont {Koseki}\ \emph {et~al.}(2025)\citenamefont {Koseki},
  \citenamefont {Aswathy},\ and\ \citenamefont
  {Rosti}}]{Koseki_Aswathy_Rosti_2025}%
  \BibitemOpen
  \bibfield  {author} {\bibinfo {author} {\bibfnamefont {M.}~\bibnamefont
  {Koseki}}, \bibinfo {author} {\bibfnamefont {M.}~\bibnamefont {Aswathy}},\
  and\ \bibinfo {author} {\bibfnamefont {M.~E.}\ \bibnamefont {Rosti}},\
  }\bibfield  {title} {\bibinfo {title} {Understanding the effect of wall
  elasticity in turbulent channel flows},\ }\href@noop {} {\bibfield  {journal}
  {\bibinfo  {journal} {Journal of Fluid Mechanics}\ }\textbf {\bibinfo
  {volume} {1021}},\ \bibinfo {pages} {A4} (\bibinfo {year}
  {2025})}\BibitemShut {NoStop}%
\bibitem [{\citenamefont {Jiménez}(2004)}]{jimenez2004turbulent}%
  \BibitemOpen
  \bibfield  {author} {\bibinfo {author} {\bibfnamefont {J.}~\bibnamefont
  {Jiménez}},\ }\bibfield  {title} {\bibinfo {title} {Turbulent flows over
  rough walls},\ }\href@noop {} {\bibfield  {journal} {\bibinfo  {journal}
  {Annu. Rev. Fluid Mech.}\ }\textbf {\bibinfo {volume} {36}},\ \bibinfo
  {pages} {173} (\bibinfo {year} {2004})}\BibitemShut {NoStop}%
\bibitem [{\citenamefont {Kadivar}\ \emph {et~al.}(2021)\citenamefont
  {Kadivar}, \citenamefont {Tormey},\ and\ \citenamefont
  {McGranaghan}}]{kadivar2021review}%
  \BibitemOpen
  \bibfield  {author} {\bibinfo {author} {\bibfnamefont {M.}~\bibnamefont
  {Kadivar}}, \bibinfo {author} {\bibfnamefont {D.}~\bibnamefont {Tormey}},\
  and\ \bibinfo {author} {\bibfnamefont {G.}~\bibnamefont {McGranaghan}},\
  }\bibfield  {title} {\bibinfo {title} {A review on turbulent flow over rough
  surfaces: Fundamentals and theories},\ }\href@noop {} {\bibfield  {journal}
  {\bibinfo  {journal} {International Journal of Thermofluids}\ }\textbf
  {\bibinfo {volume} {10}},\ \bibinfo {pages} {100077} (\bibinfo {year}
  {2021})}\BibitemShut {NoStop}%
\bibitem [{\citenamefont {Chung}\ \emph {et~al.}(2021)\citenamefont {Chung},
  \citenamefont {Hutchins}, \citenamefont {Schultz},\ and\ \citenamefont
  {Flack}}]{chung2021predicting}%
  \BibitemOpen
  \bibfield  {author} {\bibinfo {author} {\bibfnamefont {D.}~\bibnamefont
  {Chung}}, \bibinfo {author} {\bibfnamefont {N.}~\bibnamefont {Hutchins}},
  \bibinfo {author} {\bibfnamefont {M.~P.}\ \bibnamefont {Schultz}},\ and\
  \bibinfo {author} {\bibfnamefont {K.~A.}\ \bibnamefont {Flack}},\ }\bibfield
  {title} {\bibinfo {title} {Predicting the drag of rough surfaces},\
  }\href@noop {} {\bibfield  {journal} {\bibinfo  {journal} {Annual Review of
  Fluid Mechanics}\ }\textbf {\bibinfo {volume} {53}},\ \bibinfo {pages} {439}
  (\bibinfo {year} {2021})}\BibitemShut {NoStop}%
\bibitem [{\citenamefont {Kadivar}\ and\ \citenamefont
  {Garg}(2025)}]{KADIVAR2025turbulent}%
  \BibitemOpen
  \bibfield  {author} {\bibinfo {author} {\bibfnamefont {M.}~\bibnamefont
  {Kadivar}}\ and\ \bibinfo {author} {\bibfnamefont {H.}~\bibnamefont {Garg}},\
  }\bibfield  {title} {\bibinfo {title} {Turbulent heat transfer over
  roughness: a comprehensive review of theories and turbulent flow structure},\
  }\href@noop {} {\bibfield  {journal} {\bibinfo  {journal} {International
  Journal of Thermofluids}\ }\textbf {\bibinfo {volume} {26}},\ \bibinfo
  {pages} {100967} (\bibinfo {year} {2025})}\BibitemShut {NoStop}%
\bibitem [{\citenamefont {Manes}\ \emph {et~al.}(2009)\citenamefont {Manes},
  \citenamefont {Pokrajac}, \citenamefont {McEwan},\ and\ \citenamefont
  {Nikora}}]{manes2009turbulence}%
  \BibitemOpen
  \bibfield  {author} {\bibinfo {author} {\bibfnamefont {C.}~\bibnamefont
  {Manes}}, \bibinfo {author} {\bibfnamefont {D.}~\bibnamefont {Pokrajac}},
  \bibinfo {author} {\bibfnamefont {I.}~\bibnamefont {McEwan}},\ and\ \bibinfo
  {author} {\bibfnamefont {V.}~\bibnamefont {Nikora}},\ }\bibfield  {title}
  {\bibinfo {title} {Turbulence structure of open channel flows over permeable
  and impermeable beds: A comparative study},\ }\href@noop {} {\bibfield
  {journal} {\bibinfo  {journal} {Physics of Fluids}\ }\textbf {\bibinfo
  {volume} {21}} (\bibinfo {year} {2009})}\BibitemShut {NoStop}%
\bibitem [{\citenamefont {Manes}\ \emph {et~al.}(2011)\citenamefont {Manes},
  \citenamefont {Poggi},\ and\ \citenamefont {Ridolfi}}]{manes2011turbulent}%
  \BibitemOpen
  \bibfield  {author} {\bibinfo {author} {\bibfnamefont {C.}~\bibnamefont
  {Manes}}, \bibinfo {author} {\bibfnamefont {D.}~\bibnamefont {Poggi}},\ and\
  \bibinfo {author} {\bibfnamefont {L.}~\bibnamefont {Ridolfi}},\ }\bibfield
  {title} {\bibinfo {title} {Turbulent boundary layers over permeable walls:
  scaling and near-wall structure},\ }\href@noop {} {\bibfield  {journal}
  {\bibinfo  {journal} {Journal of Fluid Mechanics}\ }\textbf {\bibinfo
  {volume} {687}},\ \bibinfo {pages} {141} (\bibinfo {year}
  {2011})}\BibitemShut {NoStop}%
\bibitem [{\citenamefont {Rosti}\ \emph
  {et~al.}(2018{\natexlab{a}})\citenamefont {Rosti}, \citenamefont {Brandt},\
  and\ \citenamefont {Pinelli}}]{rosti2018turbulent}%
  \BibitemOpen
  \bibfield  {author} {\bibinfo {author} {\bibfnamefont {M.~E.}\ \bibnamefont
  {Rosti}}, \bibinfo {author} {\bibfnamefont {L.}~\bibnamefont {Brandt}},\ and\
  \bibinfo {author} {\bibfnamefont {A.}~\bibnamefont {Pinelli}},\ }\bibfield
  {title} {\bibinfo {title} {Turbulent channel flow over an anisotropic porous
  wall--drag increase and reduction},\ }\href@noop {} {\bibfield  {journal}
  {\bibinfo  {journal} {Journal of Fluid Mechanics}\ }\textbf {\bibinfo
  {volume} {842}},\ \bibinfo {pages} {381} (\bibinfo {year}
  {2018}{\natexlab{a}})}\BibitemShut {NoStop}%
\bibitem [{\citenamefont {Okazaki}\ \emph {et~al.}(2020)\citenamefont
  {Okazaki}, \citenamefont {Shimizu}, \citenamefont {Kuwata},\ and\
  \citenamefont {Suga}}]{okazaki2020turbulence}%
  \BibitemOpen
  \bibfield  {author} {\bibinfo {author} {\bibfnamefont {Y.}~\bibnamefont
  {Okazaki}}, \bibinfo {author} {\bibfnamefont {A.}~\bibnamefont {Shimizu}},
  \bibinfo {author} {\bibfnamefont {Y.}~\bibnamefont {Kuwata}},\ and\ \bibinfo
  {author} {\bibfnamefont {K.}~\bibnamefont {Suga}},\ }\bibfield  {title}
  {\bibinfo {title} {Turbulence characteristics over k-type rib roughened
  porous walls},\ }\href@noop {} {\bibfield  {journal} {\bibinfo  {journal}
  {International Journal of Heat and Fluid Flow}\ }\textbf {\bibinfo {volume}
  {82}},\ \bibinfo {pages} {108541} (\bibinfo {year} {2020})}\BibitemShut
  {NoStop}%
\bibitem [{\citenamefont {Breugem}\ \emph {et~al.}(2006)\citenamefont
  {Breugem}, \citenamefont {Boersma},\ and\ \citenamefont
  {Uittenbogaard}}]{Breugem2006The}%
  \BibitemOpen
  \bibfield  {author} {\bibinfo {author} {\bibfnamefont {W.-P.}\ \bibnamefont
  {Breugem}}, \bibinfo {author} {\bibfnamefont {B.~J.}\ \bibnamefont
  {Boersma}},\ and\ \bibinfo {author} {\bibfnamefont {R.~E.}\ \bibnamefont
  {Uittenbogaard}},\ }\bibfield  {title} {\bibinfo {title} {The influence of
  wall permeability on turbulent channel flow},\ }\href@noop {} {\bibfield
  {journal} {\bibinfo  {journal} {Journal of Fluid Mechanics}\ }\textbf
  {\bibinfo {volume} {562}},\ \bibinfo {pages} {35 } (\bibinfo {year}
  {2006})}\BibitemShut {NoStop}%
\bibitem [{\citenamefont {Suga}\ \emph {et~al.}(2010)\citenamefont {Suga},
  \citenamefont {Matsumura}, \citenamefont {Ashitaka}, \citenamefont
  {Tominaga},\ and\ \citenamefont {Kaneda}}]{SUGA2010effects}%
  \BibitemOpen
  \bibfield  {author} {\bibinfo {author} {\bibfnamefont {K.}~\bibnamefont
  {Suga}}, \bibinfo {author} {\bibfnamefont {Y.}~\bibnamefont {Matsumura}},
  \bibinfo {author} {\bibfnamefont {Y.}~\bibnamefont {Ashitaka}}, \bibinfo
  {author} {\bibfnamefont {S.}~\bibnamefont {Tominaga}},\ and\ \bibinfo
  {author} {\bibfnamefont {M.}~\bibnamefont {Kaneda}},\ }\bibfield  {title}
  {\bibinfo {title} {Effects of wall permeability on turbulence},\ }\href@noop
  {} {\bibfield  {journal} {\bibinfo  {journal} {International Journal of Heat
  and Fluid Flow}\ }\textbf {\bibinfo {volume} {31}},\ \bibinfo {pages} {974}
  (\bibinfo {year} {2010})}\BibitemShut {NoStop}%
\bibitem [{\citenamefont {Suga}\ \emph {et~al.}(2011)\citenamefont {Suga},
  \citenamefont {Mori},\ and\ \citenamefont {Kaneda}}]{Suga2011vortex}%
  \BibitemOpen
  \bibfield  {author} {\bibinfo {author} {\bibfnamefont {K.}~\bibnamefont
  {Suga}}, \bibinfo {author} {\bibfnamefont {M.}~\bibnamefont {Mori}},\ and\
  \bibinfo {author} {\bibfnamefont {M.}~\bibnamefont {Kaneda}},\ }\bibfield
  {title} {\bibinfo {title} {Vortex structure of turbulence over permeable
  walls},\ }\href@noop {} {\bibfield  {journal} {\bibinfo  {journal}
  {International Journal of Heat and Fluid Flow}\ }\textbf {\bibinfo {volume}
  {32}},\ \bibinfo {pages} {586} (\bibinfo {year} {2011})}\BibitemShut
  {NoStop}%
\bibitem [{\citenamefont {Kuwata}\ and\ \citenamefont
  {Suga}(2016)}]{kuwata_lattice_2016}%
  \BibitemOpen
  \bibfield  {author} {\bibinfo {author} {\bibfnamefont {Y.}~\bibnamefont
  {Kuwata}}\ and\ \bibinfo {author} {\bibfnamefont {K.}~\bibnamefont {Suga}},\
  }\bibfield  {title} {\bibinfo {title} {Lattice boltzmann direct numerical
  simulation of interface turbulence over porous and rough walls},\ }\href@noop
  {} {\bibfield  {journal} {\bibinfo  {journal} {International Journal of Heat
  and Fluid Flow}\ }\textbf {\bibinfo {volume} {61}},\ \bibinfo {pages} {145}
  (\bibinfo {year} {2016})}\BibitemShut {NoStop}%
\bibitem [{\citenamefont {Monti}\ \emph {et~al.}(2023)\citenamefont {Monti},
  \citenamefont {Olivieri},\ and\ \citenamefont
  {Rosti}}]{monti_olivieri_rosti_2023a}%
  \BibitemOpen
  \bibfield  {author} {\bibinfo {author} {\bibfnamefont {A.}~\bibnamefont
  {Monti}}, \bibinfo {author} {\bibfnamefont {S.}~\bibnamefont {Olivieri}},\
  and\ \bibinfo {author} {\bibfnamefont {M.~E.}\ \bibnamefont {Rosti}},\
  }\bibfield  {title} {\bibinfo {title} {Collective dynamics of dense hairy
  surfaces in turbulent flow},\ }\href@noop {} {\bibfield  {journal} {\bibinfo
  {journal} {{S}cientific {R}eports}\ }\textbf {\bibinfo {volume} {13}},\
  \bibinfo {pages} {5184} (\bibinfo {year} {2023})}\BibitemShut {NoStop}%
\bibitem [{\citenamefont {Lohrer}\ and\ \citenamefont
  {Frohlich}(2023)}]{lohrer_frohlich_2023a}%
  \BibitemOpen
  \bibfield  {author} {\bibinfo {author} {\bibfnamefont {B.}~\bibnamefont
  {Lohrer}}\ and\ \bibinfo {author} {\bibfnamefont {J.}~\bibnamefont
  {Frohlich}},\ }\bibfield  {title} {\bibinfo {title} {Large eddy simulation of
  the flow over a canopy with spanwise patches},\ }\href@noop {} {\bibfield
  {journal} {\bibinfo  {journal} {{PAMM}}\ }\textbf {\bibinfo {volume} {23}},\
  \bibinfo {pages} {e202300256} (\bibinfo {year} {2023})}\BibitemShut {NoStop}%
\bibitem [{\citenamefont {Foggi~Rota}\ \emph {et~al.}(2024)\citenamefont
  {Foggi~Rota}, \citenamefont {Monti}, \citenamefont {Olivieri},\ and\
  \citenamefont {Rosti}}]{rota2024dynamics}%
  \BibitemOpen
  \bibfield  {author} {\bibinfo {author} {\bibfnamefont {G.}~\bibnamefont
  {Foggi~Rota}}, \bibinfo {author} {\bibfnamefont {A.}~\bibnamefont {Monti}},
  \bibinfo {author} {\bibfnamefont {S.}~\bibnamefont {Olivieri}},\ and\
  \bibinfo {author} {\bibfnamefont {M.~E.}\ \bibnamefont {Rosti}},\ }\bibfield
  {title} {\bibinfo {title} {Dynamics and fluid--structure interaction in
  turbulent flows within and above flexible canopies},\ }\href@noop {}
  {\bibfield  {journal} {\bibinfo  {journal} {{J}ournal of {F}luid
  {M}echanics}\ }\textbf {\bibinfo {volume} {989}},\ \bibinfo {pages} {A11}
  (\bibinfo {year} {2024})}\BibitemShut {NoStop}%
\bibitem [{\citenamefont {Lohrer}\ and\ \citenamefont
  {Frohlich}(2025)}]{lohrer_frohlich_2025a}%
  \BibitemOpen
  \bibfield  {author} {\bibinfo {author} {\bibfnamefont {B.}~\bibnamefont
  {Lohrer}}\ and\ \bibinfo {author} {\bibfnamefont {J.}~\bibnamefont
  {Frohlich}},\ }\bibfield  {title} {\bibinfo {title} {Large-eddy simulation of
  the fluid--structure interaction in aquatic canopies consisting of highly
  flexible blades},\ }\href@noop {} {\bibfield  {journal} {\bibinfo  {journal}
  {{J}ournal of {F}luid {M}echanics}\ }\textbf {\bibinfo {volume} {1016}},\
  \bibinfo {pages} {A4} (\bibinfo {year} {2025})}\BibitemShut {NoStop}%
\bibitem [{\citenamefont {Marchioli}\ \emph {et~al.}(2025)\citenamefont
  {Marchioli}, \citenamefont {Rosti},\ and\ \citenamefont
  {Verhille}}]{marchioli_rosti_verhille_2025a}%
  \BibitemOpen
  \bibfield  {author} {\bibinfo {author} {\bibfnamefont {C.}~\bibnamefont
  {Marchioli}}, \bibinfo {author} {\bibfnamefont {M.~E.}\ \bibnamefont
  {Rosti}},\ and\ \bibinfo {author} {\bibfnamefont {G.}~\bibnamefont
  {Verhille}},\ }\bibfield  {title} {\bibinfo {title} {Flexible fibers in
  turbulence},\ }\href@noop {} {\bibfield  {journal} {\bibinfo  {journal}
  {{A}nnual {R}eview of {F}luid {M}echanics}\ }\textbf {\bibinfo {volume} {58}}
  (\bibinfo {year} {2025})}\BibitemShut {NoStop}%
\bibitem [{\citenamefont {Chandesris}\ \emph {et~al.}(2013)\citenamefont
  {Chandesris}, \citenamefont {d'Hueppe}, \citenamefont {Mathieu},
  \citenamefont {Jamet},\ and\ \citenamefont {Goyeau}}]{chandesris2013direct}%
  \BibitemOpen
  \bibfield  {author} {\bibinfo {author} {\bibfnamefont {M.}~\bibnamefont
  {Chandesris}}, \bibinfo {author} {\bibfnamefont {A.}~\bibnamefont
  {d'Hueppe}}, \bibinfo {author} {\bibfnamefont {B.}~\bibnamefont {Mathieu}},
  \bibinfo {author} {\bibfnamefont {D.}~\bibnamefont {Jamet}},\ and\ \bibinfo
  {author} {\bibfnamefont {B.}~\bibnamefont {Goyeau}},\ }\bibfield  {title}
  {\bibinfo {title} {Direct numerical simulation of turbulent heat transfer in
  a fluid-porous domain},\ }\href@noop {} {\bibfield  {journal} {\bibinfo
  {journal} {Physics of Fluids}\ }\textbf {\bibinfo {volume} {25}} (\bibinfo
  {year} {2013})}\BibitemShut {NoStop}%
\bibitem [{\citenamefont {Jouybari}\ and\ \citenamefont
  {Lundström}(2021)}]{JOUYBARI2021investigation}%
  \BibitemOpen
  \bibfield  {author} {\bibinfo {author} {\bibfnamefont {N.~F.}\ \bibnamefont
  {Jouybari}}\ and\ \bibinfo {author} {\bibfnamefont {T.~S.}\ \bibnamefont
  {Lundström}},\ }\bibfield  {title} {\bibinfo {title} {Investigation of a
  thin permeable layer effect on turbulent flow and passive scalar transport in
  a channel},\ }\href@noop {} {\bibfield  {journal} {\bibinfo  {journal}
  {Powder Technology}\ }\textbf {\bibinfo {volume} {377}},\ \bibinfo {pages}
  {115} (\bibinfo {year} {2021})}\BibitemShut {NoStop}%
\bibitem [{\citenamefont {Rosti}\ \emph
  {et~al.}(2018{\natexlab{b}})\citenamefont {Rosti}, \citenamefont {Brandt},\
  and\ \citenamefont {Pinelli}}]{rosti_brandt_pinelli_2018a}%
  \BibitemOpen
  \bibfield  {author} {\bibinfo {author} {\bibfnamefont {M.~E.}\ \bibnamefont
  {Rosti}}, \bibinfo {author} {\bibfnamefont {L.}~\bibnamefont {Brandt}},\ and\
  \bibinfo {author} {\bibfnamefont {A.}~\bibnamefont {Pinelli}},\ }\bibfield
  {title} {\bibinfo {title} {Turbulent channel flow over an anisotropic porous
  wall -- drag increase and reduction},\ }\href@noop {} {\bibfield  {journal}
  {\bibinfo  {journal} {{J}ournal of {F}luid {M}echanics}\ }\textbf {\bibinfo
  {volume} {842}},\ \bibinfo {pages} {381} (\bibinfo {year}
  {2018}{\natexlab{b}})}\BibitemShut {NoStop}%
\bibitem [{\citenamefont {Nishiyama}\ \emph {et~al.}(2020)\citenamefont
  {Nishiyama}, \citenamefont {Kuwata},\ and\ \citenamefont
  {Suga}}]{nishiyama2020direct}%
  \BibitemOpen
  \bibfield  {author} {\bibinfo {author} {\bibfnamefont {Y.}~\bibnamefont
  {Nishiyama}}, \bibinfo {author} {\bibfnamefont {Y.}~\bibnamefont {Kuwata}},\
  and\ \bibinfo {author} {\bibfnamefont {K.}~\bibnamefont {Suga}},\ }\bibfield
  {title} {\bibinfo {title} {Direct numerical simulation of turbulent heat
  transfer over fully resolved anisotropic porous structures},\ }\href@noop {}
  {\bibfield  {journal} {\bibinfo  {journal} {International Journal of Heat and
  Fluid Flow}\ }\textbf {\bibinfo {volume} {81}},\ \bibinfo {pages} {108515}
  (\bibinfo {year} {2020})}\BibitemShut {NoStop}%
\bibitem [{\citenamefont {Kuwata}\ and\ \citenamefont
  {Suga}(2017)}]{kuwata_suga_2017_direct}%
  \BibitemOpen
  \bibfield  {author} {\bibinfo {author} {\bibfnamefont {Y.}~\bibnamefont
  {Kuwata}}\ and\ \bibinfo {author} {\bibfnamefont {K.}~\bibnamefont {Suga}},\
  }\bibfield  {title} {\bibinfo {title} {Direct numerical simulation of
  turbulence over anisotropic porous media},\ }\href@noop {} {\bibfield
  {journal} {\bibinfo  {journal} {Journal of Fluid Mechanics}\ }\textbf
  {\bibinfo {volume} {831}},\ \bibinfo {pages} {41–71} (\bibinfo {year}
  {2017})}\BibitemShut {NoStop}%
\bibitem [{\citenamefont {Suga}\ \emph {et~al.}(2018)\citenamefont {Suga},
  \citenamefont {Okazaki}, \citenamefont {Ho},\ and\ \citenamefont
  {Kuwata}}]{suga2018anisotropic}%
  \BibitemOpen
  \bibfield  {author} {\bibinfo {author} {\bibfnamefont {K.}~\bibnamefont
  {Suga}}, \bibinfo {author} {\bibfnamefont {Y.}~\bibnamefont {Okazaki}},
  \bibinfo {author} {\bibfnamefont {U.}~\bibnamefont {Ho}},\ and\ \bibinfo
  {author} {\bibfnamefont {Y.}~\bibnamefont {Kuwata}},\ }\bibfield  {title}
  {\bibinfo {title} {Anisotropic wall permeability effects on turbulent channel
  flows},\ }\href@noop {} {\bibfield  {journal} {\bibinfo  {journal} {Journal
  of Fluid Mechanics}\ }\textbf {\bibinfo {volume} {855}},\ \bibinfo {pages}
  {983} (\bibinfo {year} {2018})}\BibitemShut {NoStop}%
\bibitem [{\citenamefont {Rosti}(2026)}]{rosti_2026a}%
  \BibitemOpen
  \bibfield  {author} {\bibinfo {author} {\bibfnamefont {M.~E.}\ \bibnamefont
  {Rosti}},\ }\bibfield  {title} {\bibinfo {title} {Simulating laminar and
  turbulent multiphase flows with {F}ujin},\ }\href@noop {} {\bibfield
  {journal} {\bibinfo  {journal} {{F}luid {D}ynamics {R}esearch}\ }\textbf
  {\bibinfo {volume} {58}},\ \bibinfo {pages} {021401} (\bibinfo {year}
  {2026})}\BibitemShut {NoStop}%
\bibitem [{\citenamefont {Bonet}\ and\ \citenamefont
  {Wood}(2008)}]{Bonet_Wood_2008}%
  \BibitemOpen
  \bibfield  {author} {\bibinfo {author} {\bibfnamefont {J.}~\bibnamefont
  {Bonet}}\ and\ \bibinfo {author} {\bibfnamefont {R.~D.}\ \bibnamefont
  {Wood}},\ }\href@noop {} {\emph {\bibinfo {title} {Nonlinear Continuum
  Mechanics for Finite Element Analysis}}},\ \bibinfo {edition} {2nd}\ ed.\
  (\bibinfo  {publisher} {Cambridge University Press},\ \bibinfo {year}
  {2008})\BibitemShut {NoStop}%
\bibitem [{\citenamefont {Sugiyama}\ \emph {et~al.}(2011)\citenamefont
  {Sugiyama}, \citenamefont {Ii}, \citenamefont {Takeuchi}, \citenamefont
  {Takagi},\ and\ \citenamefont {Matsumoto}}]{Sugiyama2011}%
  \BibitemOpen
  \bibfield  {author} {\bibinfo {author} {\bibfnamefont {K.}~\bibnamefont
  {Sugiyama}}, \bibinfo {author} {\bibfnamefont {S.}~\bibnamefont {Ii}},
  \bibinfo {author} {\bibfnamefont {S.}~\bibnamefont {Takeuchi}}, \bibinfo
  {author} {\bibfnamefont {S.}~\bibnamefont {Takagi}},\ and\ \bibinfo {author}
  {\bibfnamefont {Y.}~\bibnamefont {Matsumoto}},\ }\bibfield  {title} {\bibinfo
  {title} {A full eulerian finite difference approach for solving
  fluid–structure coupling problems},\ }\href@noop {} {\bibfield  {journal}
  {\bibinfo  {journal} {Journal of Computational Physics}\ }\textbf {\bibinfo
  {volume} {230}},\ \bibinfo {pages} {596} (\bibinfo {year}
  {2011})}\BibitemShut {NoStop}%
\bibitem [{\citenamefont {Rosti}\ and\ \citenamefont
  {Brandt}(2020)}]{rosti2020low}%
  \BibitemOpen
  \bibfield  {author} {\bibinfo {author} {\bibfnamefont {M.~E.}\ \bibnamefont
  {Rosti}}\ and\ \bibinfo {author} {\bibfnamefont {L.}~\bibnamefont {Brandt}},\
  }\bibfield  {title} {\bibinfo {title} {Low reynolds number turbulent flows
  over elastic walls},\ }\href@noop {} {\bibfield  {journal} {\bibinfo
  {journal} {Physics of Fluids}\ }\textbf {\bibinfo {volume} {32}} (\bibinfo
  {year} {2020})}\BibitemShut {NoStop}%
\bibitem [{\citenamefont {Yousefi}\ \emph {et~al.}(2021)\citenamefont
  {Yousefi}, \citenamefont {Ardekani}, \citenamefont {Picano},\ and\
  \citenamefont {Brandt}}]{YOUSEFI2021Regimes}%
  \BibitemOpen
  \bibfield  {author} {\bibinfo {author} {\bibfnamefont {A.}~\bibnamefont
  {Yousefi}}, \bibinfo {author} {\bibfnamefont {M.~N.}\ \bibnamefont
  {Ardekani}}, \bibinfo {author} {\bibfnamefont {F.}~\bibnamefont {Picano}},\
  and\ \bibinfo {author} {\bibfnamefont {L.}~\bibnamefont {Brandt}},\
  }\bibfield  {title} {\bibinfo {title} {Regimes of heat transfer in
  finite-size particle suspensions},\ }\href@noop {} {\bibfield  {journal}
  {\bibinfo  {journal} {International Journal of Heat and Mass Transfer}\
  }\textbf {\bibinfo {volume} {177}},\ \bibinfo {pages} {121514} (\bibinfo
  {year} {2021})}\BibitemShut {NoStop}%
\bibitem [{\citenamefont {Na}\ \emph {et~al.}(1999)\citenamefont {Na},
  \citenamefont {Papavassiliou},\ and\ \citenamefont {Hanratty}}]{NA1999use}%
  \BibitemOpen
  \bibfield  {author} {\bibinfo {author} {\bibfnamefont {Y.}~\bibnamefont
  {Na}}, \bibinfo {author} {\bibfnamefont {D.~V.}\ \bibnamefont
  {Papavassiliou}},\ and\ \bibinfo {author} {\bibfnamefont {T.~J.}\
  \bibnamefont {Hanratty}},\ }\bibfield  {title} {\bibinfo {title} {Use of
  direct numerical simulation to study the effect of prandtl number on
  temperature fields},\ }\href@noop {} {\bibfield  {journal} {\bibinfo
  {journal} {International Journal of Heat and Fluid Flow}\ }\textbf {\bibinfo
  {volume} {20}},\ \bibinfo {pages} {187} (\bibinfo {year} {1999})}\BibitemShut
  {NoStop}%
\bibitem [{\citenamefont {Zhou}\ \emph {et~al.}(2025)\citenamefont {Zhou},
  \citenamefont {Zhang}, \citenamefont {Li}, \citenamefont {Yu}, \citenamefont
  {Tian}, \citenamefont {Qiu},\ and\ \citenamefont {Su}}]{ZHOU2025dns}%
  \BibitemOpen
  \bibfield  {author} {\bibinfo {author} {\bibfnamefont {X.}~\bibnamefont
  {Zhou}}, \bibinfo {author} {\bibfnamefont {D.}~\bibnamefont {Zhang}},
  \bibinfo {author} {\bibfnamefont {X.}~\bibnamefont {Li}}, \bibinfo {author}
  {\bibfnamefont {H.}~\bibnamefont {Yu}}, \bibinfo {author} {\bibfnamefont
  {W.}~\bibnamefont {Tian}}, \bibinfo {author} {\bibfnamefont {S.}~\bibnamefont
  {Qiu}},\ and\ \bibinfo {author} {\bibfnamefont {G.}~\bibnamefont {Su}},\
  }\bibfield  {title} {\bibinfo {title} {Dns study of turbulent heat transfer
  with different prandtl numbers under constant wall-temperature difference
  condition},\ }\href@noop {} {\bibfield  {journal} {\bibinfo  {journal}
  {Progress in Nuclear Energy}\ }\textbf {\bibinfo {volume} {185}},\ \bibinfo
  {pages} {105770} (\bibinfo {year} {2025})}\BibitemShut {NoStop}%
\bibitem [{\citenamefont {Papavassiliou}\ and\ \citenamefont
  {Hanratty}(1997)}]{PAPAVASSILIOU1997transport}%
  \BibitemOpen
  \bibfield  {author} {\bibinfo {author} {\bibfnamefont {D.~V.}\ \bibnamefont
  {Papavassiliou}}\ and\ \bibinfo {author} {\bibfnamefont {T.~J.}\ \bibnamefont
  {Hanratty}},\ }\bibfield  {title} {\bibinfo {title} {Transport of a passive
  scalar in a turbulent channel flow},\ }\href@noop {} {\bibfield  {journal}
  {\bibinfo  {journal} {International Journal of Heat and Mass Transfer}\
  }\textbf {\bibinfo {volume} {40}},\ \bibinfo {pages} {1303} (\bibinfo {year}
  {1997})}\BibitemShut {NoStop}%
\bibitem [{\citenamefont {Katul}\ \emph {et~al.}(1997)\citenamefont {Katul},
  \citenamefont {Kuhn}, \citenamefont {Schieldge},\ and\ \citenamefont
  {Hsieh}}]{katul1997ejection}%
  \BibitemOpen
  \bibfield  {author} {\bibinfo {author} {\bibfnamefont {G.}~\bibnamefont
  {Katul}}, \bibinfo {author} {\bibfnamefont {G.}~\bibnamefont {Kuhn}},
  \bibinfo {author} {\bibfnamefont {J.}~\bibnamefont {Schieldge}},\ and\
  \bibinfo {author} {\bibfnamefont {C.-I.}\ \bibnamefont {Hsieh}},\ }\bibfield
  {title} {\bibinfo {title} {The ejection-sweep character of scalar fluxes in
  the unstable surface layer},\ }\href@noop {} {\bibfield  {journal} {\bibinfo
  {journal} {Boundary-Layer Meteorology}\ }\textbf {\bibinfo {volume} {83}},\
  \bibinfo {pages} {1} (\bibinfo {year} {1997})}\BibitemShut {NoStop}%
\bibitem [{\citenamefont {Kim}\ and\ \citenamefont
  {Moin}(1989)}]{kim1989transport}%
  \BibitemOpen
  \bibfield  {author} {\bibinfo {author} {\bibfnamefont {J.}~\bibnamefont
  {Kim}}\ and\ \bibinfo {author} {\bibfnamefont {P.}~\bibnamefont {Moin}},\
  }\bibfield  {title} {\bibinfo {title} {Transport of passive scalars in a
  turbulent channel flow},\ }in\ \href@noop {} {\emph {\bibinfo {booktitle}
  {Turbulent Shear Flows 6: Selected Papers from the Sixth International
  Symposium on Turbulent Shear Flows, Universit{\'e} Paul Sabatier, Toulouse,
  France, September 7--9, 1987}}}\ (\bibinfo {organization} {Springer},\
  \bibinfo {year} {1989})\ pp.\ \bibinfo {pages} {85--96}\BibitemShut {NoStop}%
\bibitem [{\citenamefont {Kasagi}\ \emph {et~al.}(1992)\citenamefont {Kasagi},
  \citenamefont {Tomita},\ and\ \citenamefont {Kuroda}}]{Kasagi1992direct}%
  \BibitemOpen
  \bibfield  {author} {\bibinfo {author} {\bibfnamefont {N.}~\bibnamefont
  {Kasagi}}, \bibinfo {author} {\bibfnamefont {Y.}~\bibnamefont {Tomita}},\
  and\ \bibinfo {author} {\bibfnamefont {A.}~\bibnamefont {Kuroda}},\
  }\bibfield  {title} {\bibinfo {title} {Direct numerical simulation of passive
  scalar field in a turbulent channel flow},\ }\href@noop {} {\bibfield
  {journal} {\bibinfo  {journal} {Journal of Heat Transfer}\ }\textbf {\bibinfo
  {volume} {114}},\ \bibinfo {pages} {598} (\bibinfo {year}
  {1992})}\BibitemShut {NoStop}%
\bibitem [{\citenamefont {Kong}\ \emph {et~al.}(2000)\citenamefont {Kong},
  \citenamefont {Choi},\ and\ \citenamefont {Lee}}]{Kong2000direct}%
  \BibitemOpen
  \bibfield  {author} {\bibinfo {author} {\bibfnamefont {H.}~\bibnamefont
  {Kong}}, \bibinfo {author} {\bibfnamefont {H.}~\bibnamefont {Choi}},\ and\
  \bibinfo {author} {\bibfnamefont {J.~S.}\ \bibnamefont {Lee}},\ }\bibfield
  {title} {\bibinfo {title} {Direct numerical simulation of turbulent thermal
  boundary layers},\ }\href@noop {} {\bibfield  {journal} {\bibinfo  {journal}
  {Physics of Fluids}\ }\textbf {\bibinfo {volume} {12}},\ \bibinfo {pages}
  {2555} (\bibinfo {year} {2000})}\BibitemShut {NoStop}%
\bibitem [{\citenamefont {Leonardi}\ \emph {et~al.}(2015)\citenamefont
  {Leonardi}, \citenamefont {Orlandi}, \citenamefont {Djenidi},\ and\
  \citenamefont {Antonia}}]{Leonardi_Orlandi_Djenidi_Antonia_2015}%
  \BibitemOpen
  \bibfield  {author} {\bibinfo {author} {\bibfnamefont {S.}~\bibnamefont
  {Leonardi}}, \bibinfo {author} {\bibfnamefont {P.}~\bibnamefont {Orlandi}},
  \bibinfo {author} {\bibfnamefont {L.}~\bibnamefont {Djenidi}},\ and\ \bibinfo
  {author} {\bibfnamefont {R.~A.}\ \bibnamefont {Antonia}},\ }\bibfield
  {title} {\bibinfo {title} {Heat transfer in a turbulent channel flow with
  square bars or circular rods on one wall},\ }\href@noop {} {\bibfield
  {journal} {\bibinfo  {journal} {Journal of Fluid Mechanics}\ }\textbf
  {\bibinfo {volume} {776}},\ \bibinfo {pages} {512–530} (\bibinfo {year}
  {2015})}\BibitemShut {NoStop}%
\bibitem [{\citenamefont {Rouhi}\ \emph {et~al.}(2022)\citenamefont {Rouhi},
  \citenamefont {Endrikat}, \citenamefont {Modesti}, \citenamefont {Sandberg},
  \citenamefont {Oda}, \citenamefont {Tanimoto}, \citenamefont {Hutchins},\
  and\ \citenamefont
  {Chung}}]{Rouhi_Endrikat_Modesti_Sandberg_Oda_Tanimoto_Hutchins_Chung_2022}%
  \BibitemOpen
  \bibfield  {author} {\bibinfo {author} {\bibfnamefont {A.}~\bibnamefont
  {Rouhi}}, \bibinfo {author} {\bibfnamefont {S.}~\bibnamefont {Endrikat}},
  \bibinfo {author} {\bibfnamefont {D.}~\bibnamefont {Modesti}}, \bibinfo
  {author} {\bibfnamefont {R.~D.}\ \bibnamefont {Sandberg}}, \bibinfo {author}
  {\bibfnamefont {T.}~\bibnamefont {Oda}}, \bibinfo {author} {\bibfnamefont
  {K.}~\bibnamefont {Tanimoto}}, \bibinfo {author} {\bibfnamefont
  {N.}~\bibnamefont {Hutchins}},\ and\ \bibinfo {author} {\bibfnamefont
  {D.}~\bibnamefont {Chung}},\ }\bibfield  {title} {\bibinfo {title}
  {Riblet-generated flow mechanisms that lead to local breaking of reynolds
  analogy},\ }\href {https://doi.org/10.1017/jfm.2022.880} {\bibfield
  {journal} {\bibinfo  {journal} {Journal of Fluid Mechanics}\ }\textbf
  {\bibinfo {volume} {951}},\ \bibinfo {pages} {A45} (\bibinfo {year}
  {2022})}\BibitemShut {NoStop}%
\bibitem [{\citenamefont {Khorasani}\ \emph {et~al.}(2025)\citenamefont
  {Khorasani}, \citenamefont {Brethouwer},\ and\ \citenamefont
  {Bagheri}}]{khorasani2025porous}%
  \BibitemOpen
  \bibfield  {author} {\bibinfo {author} {\bibfnamefont {S.~M.~H.}\
  \bibnamefont {Khorasani}}, \bibinfo {author} {\bibfnamefont {G.}~\bibnamefont
  {Brethouwer}},\ and\ \bibinfo {author} {\bibfnamefont {S.}~\bibnamefont
  {Bagheri}},\ }\href {https://arxiv.org/abs/2507.08962} {\bibinfo {title}
  {Turbulent heat transfer in open-channel flows with a thermally-conductive
  porous wall}} (\bibinfo {year} {2025}),\ \Eprint
  {https://arxiv.org/abs/2507.08962} {arXiv:2507.08962 [physics.flu-dyn]}
  \BibitemShut {NoStop}%
\end{thebibliography}
\providecommand{\noopsort}[1]{}\providecommand{\singleletter}[1]{#1}%

\end{document}